\documentclass[preprintnumbers, floatfix,letterpaper,aps,prd,epsfig,nofootinbib,
longbibliography,
twocolumn
]{revtex4-1}
\usepackage{bm,graphicx,dcolumn,epstopdf,epsf, latexsym,mathbbol, amssymb,amsmath,color,slashed, mathrsfs,mathcomp,simplewick}
\usepackage[center]{subfigure}
\usepackage{multirow}
\usepackage{makecell}
\usepackage{ytableau}
\usepackage{booktabs}
\usepackage[colorlinks,linkcolor=blue,citecolor=blue,urlcolor=blue]{hyperref}

\begin{document}
\allowdisplaybreaks
 \newcommand{\bq}{\begin{equation}}
 \newcommand{\eq}{\end{equation}}
 \newcommand{\bqn}{\begin{eqnarray}}
 \newcommand{\eqn}{\end{eqnarray}}
 \newcommand{\nb}{\nonumber}
 \newcommand{\lb}{\label}
 \newcommand{\f}{\frac}
 \newcommand{\p}{\partial}
\newcommand{\PRL}{Phys. Rev. Lett.}
\newcommand{\PLB}{Phys. Lett. B}
\newcommand{\PRD}{Phys. Rev. D}
\newcommand{\CQG}{Class. Quantum Grav.}
\newcommand{\JCAP}{J. Cosmol. Astropart. Phys.}
\newcommand{\JHEP}{J. High. Energy. Phys.}
\newcommand{\bea}{\begin{eqnarray}}
\newcommand{\ena}{\end{eqnarray}}
\newcommand{\beqa}{\begin{eqnarray}}
\newcommand{\eeqa}{\end{eqnarray}}
\newcommand{\red}{\textcolor{red}}

\newlength\scratchlength
\newcommand\s[2]{
  \settoheight\scratchlength{\mathstrut}%
  \scratchlength=\number\numexpr\number#1-1\relax\scratchlength
  \lower.5\scratchlength\hbox{\scalebox{1}[#1]{$#2$}}%
}


\title{Modified gravitational wave propagations in linearized gravity with Lorentz and diffeomorphism violations and their gravitational wave constraints}

\author{Qiang Wang${}^{a, b}$}
\email{qiangwang@zjut.edu.cn}

\author{Jian-Ming Yan${}^{a, b}$}
\email{yanjm@zjut.edu.cn}

\author{Tao Zhu${}^{a, b}$}
\email{Corresponding author: zhut05@zjut.edu.cn}

\author{Wen Zhao${}^{c, d}$}
\email{wzhao7@ustc.edu.cn}

\affiliation{${}^{a}$Institute for Theoretical Physics \& Cosmology, Zhejiang University of Technology, Hangzhou, 310023, China\\
${}^{b}$ United Center for Gravitational Wave Physics (UCGWP),  Zhejiang University of Technology, Hangzhou, 310023, China\\
 ${}^{c}$CAS Key Laboratory for Research in Galaxies and Cosmology, Department of Astronomy, University of Science and Technology of China, Hefei 230026, China \\
${}^{d}$School of Astronomy and Space Sciences, University of Science and Technology of China, Hefei, 230026, China}

\date{\today}

\begin{abstract}

The standard model extension (SME) is an effective field theory framework that can be used to study the possible violations of Lorentz symmetry and diffeomorphism invariance in the gravitational interaction. In this paper, we explore both the Lorentz- and diffeomorphism-violating effects on the propagations of gravitational waves in the SME's linearized gravity. It is shown that the violations of Lorentz symmetry and diffeomorphism invariance modify the conventional linear dispersion relation of gravitational waves, leading to anisotropy, birefringence, and dispersion effects in the propagation of gravitational waves. With these modified dispersion relations, we then calculate the dephasing effects due to the Lorentz and diffeomorphism violations in the waveforms of gravitational waves produced by the coalescence of compact binaries. With the distorted waveforms, we perform full Bayesian inference with the help of the open source software \texttt{BILBY} on the gravitational wave events of the compact binary mergers in the LIGO-Virgo-KAGRA catalogs GWTC-3. We consider the effects from the operators with the lowest mass dimension $d=2$ and $d=3$ due to the Lorentz and diffeomorphism violations in the linearized gravity.  No signature of Lorentz and diffeomorphism violations arsing from the SME's linearized gravity are found for most GW events, which allows us to give a $90\%$ confidence interval for each Lorentz- and diffeomorphism-violating coefficient. 

\end{abstract}


\maketitle

\section{Introduction}
\renewcommand{\theequation}{1.\arabic{equation}} \setcounter{equation}{0}

General relativity (GR) stands as the preeminent theory of gravity, having undergone rigorous experimental validation across a diverse range of scales with remarkable precision \cite{Berti:2015itd, Will:1997bb, Hoyle:2000cv, Jain:2010ka, Koyama:2015vza, Clifton:2011jh, Stairs:2003eg, Manchester:2015mda, Wex:2014nva, Kramer:2016kwa}. Despite its empirical success, GR encounters significant challenges related to the theoretical singularities and issues of quantization, as well as the unresolved phenomena of dark matter and dark energy. Conversely, several candidate theories of quantum gravity, including string theory \cite{Kostelecky:1988zi, Kostelecky:1991ak}, loop quantum gravity \cite{Gambini:1998it}, and brane-world scenarios \cite{Burgess:2002tb}, propose frameworks wherein Lorentz and diffeomorphism invariances in the gravitational sector may be spontaneously violated. This potential breakdown of Lorentz and diffeomorphism invariances presents intriguing avenues for addressing some of the limitations of GR and advancing our understanding of fundamental physics.

A common method to investigate potential violations of Lorentz and diffeomorphism symmetries in gravity is through the approach of effective field theory. The Standard-Model Extension (SME) offers a comprehensive framework for examining deviations from Lorentz and diffeomorphism invariance  \cite{Colladay:1996iz, Kostelecky:2003fs}. Within this approach, any components that might disrupt Lorentz or/and diffeomorphism invariance can be systematically included in the Lagrangian. Over recent decades, the SME has been widely applied to test Lorentz invariance in the matter sector. In the context of gravity, studies leveraging the SME to evaluate the Lorentz violations have encompassed a range of techniques including lunar laser ranging \cite{Battat:2007uh,Bourgoin:2017fpo}, atom interferometry \cite{Muller:2007es}, cosmic ray observations \cite{Kostelecky:2015dpa}, precision pulsar timing \cite{Shao:2014oha,Shao:2014bfa,Jennings:2015vma,Shao:2018vul,Shao:2019cyt,Shao:2019tle}, planetary orbital analyses \cite{Hees:2015mga}, and superconducting gravimeters \cite{Flowers:2016ctv}. Studies in the gravity domain are primarily concerned with the interaction between gravity and matter \cite{Kostelecky:2010ze}. Our focus, however, is on the Lorentz and diffeomorphism-violating effects on gravitational wave (GW) propagation in the pure gravity sector within a linear approximation \cite{Bailey:2006fd, Ferrari:2006gs, Kostelecky:2017zob}.

In linearized Lorentz-violating gravity, the possible Lorentz-violating terms involving quadratic metric variations in the Lagrangian can be classified by their mass dimension, \(d\). These terms are further divided into two categories: diffeomorphism-invariant terms and diffeomorphism-violating terms, depending on whether they remain invariant under the gauge transformation \(h_{\mu \nu} \rightarrow h_{\mu \nu} + \partial_\mu \xi_\nu + \partial_\nu \xi_\mu\). Here, \(h_{\mu\nu}\) represents metric perturbations in Minkowski space-time, and \(\xi_\mu\) is an arbitrary small vector field. Diffeomorphism invariance requires that \(d \geq 4\) \cite{Kostelecky:2016kfm, Kostelecky:2017zob}. The effects of Lorentz violations on gravitational waves (GWs) and their observational constraints have been extensively studied for the diffeomorphism-invariant case; 
for example, \cite{Hou:2024xbv, Kostelecky:2016kfm, Kostelecky:2017zob, Dong:2023nau, Ault:2023sld, Zhu:2023rrx, Gong:2023ffb, Haegel:2022ymk, Niu:2022yhr, ONeal-Ault:2021uwu, Shao:2020shv, Mewes:2019dhj} and references therein. In contrast, for the diffeomorphism-violating case, the Lorentz-violating terms in the Lagrangian can have mass dimensions \(d = 2\) or/and \(d = 3\) \cite{Kostelecky:2017zob}.

An exhaustive classification of Lorentz-violating Lagrangians with quadratic metric variations, whether diffeomorphism-invariant or -violating, is provided in \cite{Kostelecky:2017zob}. These terms modify GW dispersion relations, leading to effects such as anisotropy, birefringence, and dispersion, which alter GW waveforms. Waveform modifications for the diffeomorphism-invariant case are discussed in \cite{Mewes:2019dhj} and can be analyzed using Bayesian inference to compare observed GW signals with theoretical models, constraining the Lorentz-violating coefficients in the SME's linearized gravity sector. A parametrized framework for describing symmetry-breaking effects on GWs is presented in \cite{Zhu:2023rrx}, which has been used for calculating the symmetry-breaking effects on the primordial GWs \cite{Li:2024fxy}. In this paper, we investigate both Lorentz-violating and diffeomorphism-violating effects on GW propagation and calculate the resulting dephasing in waveforms from compact binary coalescences.

The direct detection of GWs from compact binary coalescences by the LIGO-Virgo-KAGRA (LVK) Collaboration has ushered in a new era of gravitational physics \cite{LIGOScientific:2016aoc, KAGRA:2021vkt, LIGOScientific:2020ibl, LIGOScientific:2018mvr, LIGOScientific:2017vwq, LIGOScientific:2016emj, LIGOScientific:2016vlm, LIGOScientific:2016vbw}. These GWs carry crucial information about the local spacetime properties of compact binaries, enabling tests of fundamental gravitational symmetries. Numerous studies have tested Lorentz and parity symmetries using data from LVK GW events \cite{Kostelecky:2016kfm, Wang:2020cub, Zhao:2019xmm, Zhao:2022pun, Wang:2020pgu, Zhu:2022uoq, Zhu:2022dfq, Niu:2022yhr, Qiao:2019hkz, Wang:2021gqm, Zhao:2019szi, Chen:2022wtz, Li:2022grj, Haegel:2022ymk, Qiao:2021fwi, ONeal-Ault:2021uwu, Sulantay:2022sag, Wang:2020pgu, Wang:2017igw, LIGOScientific:2019fpa, LIGOScientific:2020tif, LIGOScientific:2021sio}; see \cite{Qiao:2022mln} for a recent review. While previous tests in Lorentz-violating linearized gravity for diffeomorphism-invariant cases focused on terms with mass dimension \(d \geq 4\), this paper examines the effects of diffeomorphism violations induced by Lorentz-violating terms with \(d = 2\) and \(d = 3\), along with their observational constraints from LVK GW events. Using the SME framework for Lorentz-violating linearized gravity, we perform Bayesian inference with modified waveforms incorporating diffeomorphism-violating effects on GW events from the GWTC-3 catalog. We find no significant evidence of diffeomorphism violations in most GW data and provide 90\% confidence intervals for each diffeomorphism-violating coefficient.

This paper is organized as follows. In the next section, we provide a brief overview of GW propagation in the SME framework, incorporating both Lorentz and diffeomorphism violations and their modified dispersion relations. Section \ref{sec3} examines phase modifications to GW waveforms caused by Lorentz-violating and diffeomorphism-violating coefficients in the SME. In Section \ref{sec4}, we describe the matched-filter analysis within Bayesian inference, and in Section \ref{sec5}, we present constraints on each diffeomorphism-violating coefficient using data from GW events in the GWTC-3 catalog. Finally, the conclusions and summary are provided in Section \ref{sec6}.

Throughout this paper, we adopt the metric convention $(-,+,+,+)$, with Greek indices $(\mu, \nu, \dots)$ running over $0,1,2,3$ and Latin indices $(i, j, k)$ running over $1,2,3$. Natural units are used, setting $\hbar = c = 1$. 

\section{Gravitational wave propagations in the linear gravity of SME \label{sec2}}
\renewcommand{\theequation}{2.\arabic{equation}} \setcounter{equation}{0}

In this section, we present a brief introduction to the GWs in the linearized gravity sector of the SME and the associated modified dispersion relation of GWs due to the effects of the Lorentz and diffeomorphism violations. 

\subsection{Linearized gravity with Lorentz and diffeomorphism violations}

The quadratic Lagrangian density for GWs  in the  linearized gravity sector of the SME is given by \cite{Kostelecky:2017zob}
\bqn \lb{calL}
\mathcal{L} &=& \frac{1}{4}\epsilon^{\mu \rho \alpha \kappa} \epsilon^{\nu \sigma \beta \lambda} \eta_{\kappa \lambda} h_{\mu\nu} \partial_\alpha \partial_\beta h_{\rho \sigma} \nb\\
&&+ \frac{1}{4}h_{\mu\nu} \sum_{{\cal K}, d}  \hat{\cal K}^{(d)\mu\nu\rho\sigma}h_{\rho \sigma},
\eqn
where one expands the metric $g_{\mu\nu}$ of the spacetime in the form of $g_{\mu \nu} = \eta_{\mu\nu} + h_{\mu\nu}$ with $\eta_{\mu\nu}$ being the constant Minkowski metric, and $\epsilon^{\mu\rho \alpha \kappa}$ is the Levi-Civita tensor. The first term in the above expression represents the quadratic approximation to the Lagrangian density for the Einstein-Hilbert action, while the second term which consists of operators $\hat{\cal K}^{(d)\mu\nu\rho\sigma}$ denotes the modifications due to the Lorentz and diffeomorphism violations. The operator $\hat{\cal K}^{(d)\mu\nu\rho\sigma}$ is the product of a coefficient ${\cal K}^{(d)\mu\nu\rho\sigma {\alpha _1 \alpha _2 \cdots \alpha _{d-2}}}$ with $d-2$ derivatives {$\partial_{\alpha _1 } \partial_{\alpha _2} \cdots \partial_{\alpha _{d-2}}$}, i.e., 
\bqn
\hat{\cal K}^{(d)\mu\nu\rho\sigma} = {\cal K}^{(d)\mu\nu\rho\sigma \alpha _1 \alpha _2 \cdots \alpha _{d-2}}{\partial_{\alpha _1 } \partial_{\alpha _2} \cdots \partial_{\alpha _{d-2}}}.
\eqn
The coefficient ${\cal K}^{(d)\mu\nu\rho\sigma \alpha _1 \alpha _2 \cdots \alpha _{d-2}}$ has mass dimension $4-d$ and are assumed small and constant over the scales relevant for the gravitational phenomenon considered in this paper. As pointed out in \cite{Kostelecky:2017zob}, to contribute non-trivially to the equation of motion, the $\hat{\cal K}^{(d)\mu\nu\rho\sigma}$ has to satisfy the requirement $\hat{\cal K}^{(d) (\mu\nu) (\rho\sigma)}  \neq \pm \hat{\cal K}^{(d)  (\rho\sigma)( \mu\nu)}$, where $\pm$ corresponds to odd $d$ and even $d$ respectively. 

According to the the symmetries of the coefficient $ {\cal K}^{(d)\mu\nu\rho\sigma {\alpha _1 \alpha _2 \cdots \alpha _{d-2}}}$ under the permutations of its indices, $\hat{\cal K}^{(d)\mu\nu\rho\sigma}$ can be divided into three different types,
\bqn
\hat{\cal K}^{(d)\mu\nu\rho\sigma} = {\hat s}^{(d)\mu\rho\nu\sigma}+{\hat q}^{(d)\mu\rho\nu\sigma}+{\hat k}^{(d)\mu\rho\nu\sigma}.
\eqn
These three types of operators have different symmetries in their indices. Specifically, $\hat{s}^{(d)\mu \rho \nu \sigma}$ is anti-symmetric in both ``$\mu \rho$" and ``$\nu\sigma$", $\hat{q}^{(d)\mu \rho \nu \sigma}$ is anti-symmetric in ``$\mu\rho$" and symmetric in ``$\nu\sigma$", and $\hat{k}^{(d)\mu \rho \nu \sigma}$ is totally symmetric. One can further decompose each type of operator into several irreducible pieces and explore their proerties on the gravitational wave propagations. It is shown in \cite{Kostelecky:2017zob} such a decomposition leads to 14 independent classess of operators in total, as preented in Table \ref{young}, see also Table I of ref. \cite{Kostelecky:2017zob}.  

The $s$-type operators ${\hat s}^{(d)\mu\rho\nu\sigma}$ are CPT even, denoting that these operators are invariant under the combined symmetry of change conjugation, parity transformation, and time reversal (CPT). They consist of three irreducible pieces in the decomposion, 
\bqn
\hat{s}^{{(d)\mu\rho\nu\sigma}} = \hat{s}^{(d, 0)\mu\rho\nu\sigma} + \hat{s}^{(d, 1)\mu\rho\nu\sigma} +  \hat{s}^{(d, 2)\mu\rho\nu\sigma},
\eqn
with
\bqn\lb{sd}
{\hat{s}}^{(d,0)\mu \rho \nu \sigma} &=& {s}^{(d,0) \mu \rho \alpha_{1} \nu \sigma \alpha_{2} \alpha_3 \ldots \alpha_{d-2}} \partial_{\alpha_{1}} \ldots \partial_{\alpha_{d-2}}, \lb{sd0}\nb\\
{\hat{s}}^{(d,1)\mu \rho \nu \sigma} &=& {s}^{(d,1) \mu \rho \nu \sigma \alpha_{1}  \ldots \alpha_{d-2}} \partial_{\alpha_{1}} \ldots \partial_{\alpha_{d-2}}, \lb{sd1}\nb\\
{\hat{s}}^{(d,2)\mu \rho \nu \sigma} &=& {s}^{(d,2) \mu \rho \alpha_{1} \nu \sigma \alpha_{2} \alpha_3 \ldots \alpha_{d-2}} \partial_{\alpha_{1}} \ldots \partial_{\alpha_{d-2}}. \lb{sd2}
\eqn
The $q$-type operators $\hat{q}^{(d)\mu\rho\nu\sigma}$ are CPT odd, denoting that they change sign under the symmetry of CPT. They consist of six irreducible pieces in the decomposion,
\bqn
\hat{q}^{(d)\mu\rho\nu\sigma} &=& \hat{q}^{(d, 0)\mu\rho\nu\sigma}+\hat{q}^{(d, 1)\mu\rho\nu\sigma}+\hat{q}^{(d, 2)\mu\rho\nu\sigma}\nb\\
&&+\hat{q}^{(d, 3)\mu\rho\nu\sigma}+\hat{q}^{(d, 4)\mu\rho\nu\sigma}+\hat{q}^{(d, 5)\mu\rho\nu\sigma},\nb\\
\eqn
where
\bqn\lb{qd}
\hat{q}^{(d, 0)\mu\rho\nu\sigma}&=&\hat{q}^{(d, 0)\mu\rho\alpha_1 \nu \alpha_2 \sigma \alpha_3 \alpha_4 \ldots \alpha_{d-2}} \partial_{\alpha_{1}} \ldots \partial_{\alpha_{d-2}}, \nb\\
\hat{q}^{(d, 1)\mu\rho\nu\sigma}&=&\hat{q}^{(d, 1)\mu\rho \nu  \sigma \alpha_1 \alpha_2 \ldots \alpha_{d-2}} \partial_{\alpha_{1}} \ldots \partial_{\alpha_{d-2}}, \nb\\
\hat{q}^{(d, 2)\mu\rho\nu\sigma}&=&\hat{q}^{(d, 2)\mu\rho \nu \alpha_1 \sigma \alpha_2  \ldots \alpha_{d-2}} \partial_{\alpha_{1}} \ldots \partial_{\alpha_{d-2}}, \nb\\
\hat{q}^{(d, 3)\mu\rho\nu\sigma}&=&\hat{q}^{(d, 3)\mu\rho\alpha_1 \nu  \sigma \alpha_2 \ldots \alpha_{d-2}} \partial_{\alpha_{1}} \ldots \partial_{\alpha_{d-2}}, \nb\\
\hat{q}^{(d, 4)\mu\rho\nu\sigma}&=&\hat{q}^{(d, 4)\mu\rho\nu \alpha_1 \sigma \alpha_2 \alpha_3 \alpha_4 \ldots \alpha_{d-2}} \partial_{\alpha_{1}} \ldots \partial_{\alpha_{d-2}}, \nb\\
\hat{q}^{(d, 5)\mu\rho\nu\sigma}&=&\hat{q}^{(d, 5)\mu\rho \alpha_1 \nu \alpha_2 \sigma \alpha_3 \alpha_4 \ldots \alpha_{d-2}} \partial_{\alpha_{1}} \ldots \partial_{\alpha_{d-2}}.\nb\\
\eqn
And the $k$-type operators are CPT even and consist of five irreducible pieces,
\bqn
\hat{k}^{(d)\mu\nu\rho\sigma} &=& \hat{k}^{(d, 0)\mu\nu\rho\sigma} +\hat{k}^{(d, 1)\mu\nu\rho\sigma} +\hat{k}^{(d, 2)\mu\nu\rho\sigma} \nb\\
&&+\hat{k}^{(d, 3)\mu\nu\rho\sigma} +\hat{k}^{(d, 4)\mu\nu\rho\sigma} ,
\eqn
where
\bqn\lb{kd}
\hat{k}^{(d, 0)\mu\nu\rho\sigma} &=& \hat{k}^{(d, 0)\mu \alpha_1 \nu \alpha_2 \rho \alpha_3 \sigma \alpha_4 \alpha_5 \ldots \alpha_{d-2}}  \partial_{\alpha_{1}} \ldots \partial_{\alpha_{d-2}}, \nb\\
\hat{k}^{(d, 1)\mu\nu\rho\sigma} &=& \hat{k}^{(d, 1)\mu \nu  \rho \sigma \alpha_1 \ldots \alpha_{d-2}}  \partial_{\alpha_{1}} \ldots \partial_{\alpha_{d-2}}, \nb\\
\hat{k}^{(d, 2)\mu\nu\rho\sigma} &=& \hat{k}^{(d, 2)\mu \alpha_1 \nu  \rho  \sigma \alpha_1 \alpha_2 \ldots \alpha_{d-2}}  \partial_{\alpha_{1}} \ldots \partial_{\alpha_{d-2}}, \nb\\
\hat{k}^{(d, 3)\mu\nu\rho\sigma} &=& \hat{k}^{(d, 3)\mu \alpha_1 \nu \alpha_2 \rho  \sigma \alpha_3 \alpha_5 \ldots \alpha_{d-2}}  \partial_{\alpha_{1}} \ldots \partial_{\alpha_{d-2}}, \nb\\
\hat{k}^{(d, 4)\mu\nu\rho\sigma} &=& \hat{k}^{(d, 4)\mu \alpha_1 \nu \alpha_2 \rho \alpha_3 \sigma \alpha_4 \alpha_5 \ldots \alpha_{d-2}}  \partial_{\alpha_{1}} \ldots \partial_{\alpha_{d-2}}. \nb\\
\eqn
The properties of the above 14 coefficients are summarized in Table.~\ref{young}, see also Table I in \cite{Kostelecky:2017zob} for more detailed properties of these coefficients. These 14 classes thus characterize all possible phenomenological effects in linearized gravity, affecting the propagating properties of gravitational waves. 

If we restrict the theory to be diffeomorphism invariance, the quadratic action $S \sim \int d^4 x \mathcal{L}$ of the linearized gravity should be invariant under the the gauge transformation \(h_{\mu \nu} \rightarrow h_{\mu \nu} +\partial_\mu \xi_\nu + \partial_\nu \xi_\mu\), which requires the condition $\hat{\cal K}^{(d) (\mu\nu)(\rho\sigma)}\partial_\nu  = \pm \hat{\cal K}^{(d)  (\rho\sigma)(\mu\nu)}\partial_\nu$ holds. With this condition, the operator $\hat{\cal K}^{(d)  \mu\nu \rho\sigma}$ can only be decomposed into three independent classes \cite{Kostelecky:2017zob}, which are represents by ${\hat{s}}^{(d,0)\mu \rho \nu \sigma}$, ${\hat{q}}^{(d,0)\mu \rho \nu \sigma}$, and ${\hat{k}}^{(d,0)\mu \rho \nu \sigma}$, respectively. It is obvious that the diffeomorphism-invariance case only allows the Lorentz-violating operators with mass dimension $d\geq 4$, while the diffeomorphism-violating case allows operators with mass dimension $d\geq 2$.

\begin{table*}
\caption{Properties of the coefficients $\hat{\cal K}^{(d)\mu\nu\rho\sigma \alpha_1\ldots \alpha_{d-2}}$ in each irreducible class. A similar table can also be found in Table.~I of ref.~\cite{Kostelecky:2017zob}.}
\label{young}
\begin{ruledtabular}
    \begin{tabular}{llcc}
    Coefficients $\hat{\cal K}^{(d)\mu\nu\rho\sigma \alpha_1\ldots \alpha_{d-2}}$ & Young Tableau & $d$ &  number \\
    \hline
    \;\;\\
  $\hat{s}^{(d,0) \mu \rho \alpha_{1} \nu \sigma \alpha_{2} \alpha_3 \ldots \alpha_{d-2}} $ 
  &
\ytableausetup{centertableaux,boxsize=2.5em}
\scalebox{1}[.5]{
\begin{ytableau}
\s{2}{\mu} & \s{2}{\nu} & \s{2}{\alpha_3} & \none[\cdots] & \s{2}{\alpha_{d-2}}\\
\s{2}{\rho} & \s{2}{\sigma} \\
\s{2}{\alpha_1} &\s{2}{ \alpha_2}
\end{ytableau} }
&
 even $d \geq 4$ 
 &
$(d-3)(d-2)(d+1)$\\
\;\;\\
 $\hat{s}^{(d,1) \mu \rho \nu \sigma \alpha_{1}  \ldots \alpha_{d-2}} $ 
 &
 \ytableausetup{centertableaux,boxsize=2.5em}
 \scalebox{1}[.5]{
\begin{ytableau}
\s{2}{\mu} & \s{2}{\nu} & \s{2}{\alpha_1} & \none[\cdots] &\s{2}{\alpha_{d-2}}\\
\s{2}{\rho} & \s{2}{\sigma} 
\end{ytableau} }
&
even $d \geq 2$ 
 &
$(d-1)(d+2)(d+3)$\\
\;\;\\
 $\hat{s}^{(d,2) \mu \rho \alpha_{1} \nu \sigma \alpha_{2} \alpha_3 \ldots \alpha_{d-2}} $ 
 &
 \ytableausetup{centertableaux,boxsize=2.5em}
 \scalebox{1}[.5]{
\begin{ytableau}
\s{2}{\mu} & \s{2}{\nu} & \s{2}{\alpha_2} & \s{2}{\alpha_3} &  \none[\cdots] & \s{2}{\alpha_{d-2}}\\
\s{2}{\rho} & \s{2}{\sigma} \\
\s{2}{\alpha_1} 
\end{ytableau} }
&
even $d \geq 4$ 
 &
$\frac{4}{3}(d-2)d(d+2)$\\
\;\;\\
\hline
 \;\;\\
  $\hat{q}^{(d, 0)\mu\rho\alpha_1 \nu \alpha_2 \sigma \alpha_3 \alpha_4 \ldots \alpha_{d-2}} $ 
  &
\ytableausetup{centertableaux,boxsize=2.5em}
\scalebox{1}[.5]{
\begin{ytableau}
\s{2}{\mu} & \s{2}{\nu} & \s{2}{\sigma}  & \s{2}{\alpha_4} &  \none[\cdots] & \s{2}{\alpha_{d-2}}\\
\s{2}{\rho} & \s{2}{\alpha_2} & \s{2}{\alpha_3} \\
\s{2}{\alpha_1} 
\end{ytableau}  }
&
 odd $d \geq 5$ 
 &
$\frac{5}{2}(d-4)(d-1)(d+1)$\\
\;\;\\
  $\hat{q}^{(d, 1)\mu\rho \nu  \sigma \alpha_1 \alpha_2 \ldots \alpha_{d-2}} $ 
  &
\ytableausetup{centertableaux,boxsize=2.5em}
\scalebox{1}[.5]{
\begin{ytableau}
\s{2}{\mu} &\s{2}{ \nu} & \s{2}{\sigma} & \s{2}{\alpha_1} & \s{2}{\alpha_2}  & \none[\cdots] & \s{2}{\alpha_{d-2}}\\
\s{2}{\rho} 
\end{ytableau} }
&
 odd $d \geq 3$ 
 &
$\frac{1}{2}(d-3)(d+4)(d+1)$\\
\;\;\\
  $\hat{q}^{(d, 2)\mu\rho \nu \alpha_1 \sigma \alpha_2  \ldots \alpha_{d-2}}  $ 
  &
\ytableausetup{centertableaux,boxsize=2.5em}
\scalebox{1}[.5]{
\begin{ytableau}
\s{2}{\mu }& \s{2}{\nu} & \s{2}{\sigma} & \s{2}{\alpha_2} & \none[\cdots] & \s{2}{\alpha_{d-2}}\\
\s{2}{\rho} & \s{2}{\alpha_1} \\
\end{ytableau}  }
&
 odd $d \geq 3$ 
 &
$(d-1)(d+2)(d+3)$\\
\;\;\\
  $\hat{q}^{(d, 3)\mu\rho\alpha_1 \nu  \sigma \alpha_2 \ldots \alpha_{d-2}}$ 
  &
\ytableausetup{centertableaux,boxsize=2.5em}
\scalebox{1}[.5]{
\begin{ytableau}
\s{2}{\mu }& \s{2}{\nu} & \s{2}{\sigma} & \s{2}{\alpha_2} & \none[\cdots] & \s{2}{\alpha_{d-2}}\\
\s{2}{\rho}  \\
\s{2}{\alpha_1} 
\end{ytableau}  }
&
 odd $d \geq 3$ 
 &
$\frac{1}{2 }d (d+3)(d+1)$\\
\;\;\\
  $\hat{q}^{(d, 4)\mu\rho\nu \alpha_1 \sigma \alpha_2 \alpha_3 \alpha_4 \ldots \alpha_{d-2}}$ 
  &
\ytableausetup{centertableaux,boxsize=2.5em}
\scalebox{1}[.5]{
\begin{ytableau}
\s{2}{\mu} & \s{2}{\nu} & \s{2}{ \sigma}  &\s{2}{\alpha_3} & \s{2}{\alpha_4} & \none[\cdots] & \s{2}{\alpha_{d-2}}\\
\s{2}{\rho} &\s{2}{ \alpha_1} & \s{2}{\alpha_2}
\end{ytableau} }
&
 odd $d \geq 5$ 
 &
$\frac{5}{3}(d-3)(d+2)(d+1)$\\
\;\;\\
  $\hat{q}^{(d, 5)\mu\rho \alpha_1 \nu \alpha_2 \sigma \alpha_3 \alpha_4 \ldots \alpha_{d-2}}$ 
  &
\ytableausetup{centertableaux,boxsize=2.5em}
\scalebox{1}[.5]{
\begin{ytableau}
\s{2}{\mu} & \s{2}{\nu} & \s{2}{ \sigma}  &\s{2}{\alpha_3} & \s{2}{\alpha_4} & \none[\cdots] & \s{2}{\alpha_{d-2}}\\
\s{2}{\rho}& \s{2}{\alpha_2}\\
 \s{2}{ \alpha_1} 
\end{ytableau} }
&
 odd $d \geq 5$ 
 &
$\frac{4}{3}(d+2)d(d-2)$\\
 \;\;\\
 \hline
  \;\;\\
  $ \hat{k}^{(d, 0)\mu \alpha_1 \nu \alpha_2 \rho \alpha_3 \sigma \alpha_4 \alpha_5 \ldots \alpha_{d-2}}  $ 
  &
\ytableausetup{centertableaux,boxsize=2.5em}
\scalebox{1}[.5]{
\begin{ytableau}
\s{2}{\mu} & \s{2}{\nu} & \s{2}{ \rho}  &\s{2}{\sigma} & \s{2}{\alpha_5} & \none[\cdots] & \s{2}{\alpha_{d-2}}\\
\s{2}{\alpha_1}& \s{2}{\alpha_2} & \s{2}{\alpha_3} & \s{2}{\alpha_4}
\end{ytableau} }
&
 even $d \geq 6$ 
 &
$\frac{5}{2}(d-5)d (d+1)$\\
\;\;\\
  $ \hat{k}^{(d, 1)\mu \nu  \rho \sigma \alpha_1 \ldots \alpha_{d-2}}  $ 
  &
\ytableausetup{centertableaux,boxsize=2.5em}
\scalebox{1}[.5]{
\begin{ytableau}
\s{2}{\mu} & \s{2}{\nu} & \s{2}{ \rho}  &\s{2}{\sigma} & \s{2}{\alpha_1} & \none[\cdots] & \s{2}{\alpha_{d-2}}\\
\end{ytableau} }
&
 even $d \geq 2$ 
 &
$\frac{1}{6}(d+3)(d+4)(d+5)$\\
\;\;\\
  $ \hat{k}^{(d, 2)\mu \alpha_1 \nu  \rho  \sigma \alpha_1 \alpha_2 \ldots \alpha_{d-2}} $ 
  &
\ytableausetup{centertableaux,boxsize=2.5em}
\scalebox{1}[.5]{
\begin{ytableau}
\s{2}{\mu} & \s{2}{\nu} & \s{2}{ \rho}  &\s{2}{\sigma} & \s{2}{\alpha_2} & \s{2}{\alpha_3} & \none[\cdots] & \s{2}{\alpha_{d-2}}\\
\s{2}{\alpha_1}
\end{ytableau}}
&
 even $d \geq 4$ 
 &
$\frac{1}{2}(d+1)(d+3)(d+4)$\\
\;\;\\
  $ \hat{k}^{(d, 3)\mu \alpha_1 \nu \alpha_2 \rho  \sigma \alpha_3 \alpha_5 \ldots \alpha_{d-2}} $ 
  &
\ytableausetup{centertableaux,boxsize=2.5em}
\scalebox{1}[.5]{
\begin{ytableau}
\s{2}{\mu} & \s{2}{\nu} & \s{2}{ \rho}  &\s{2}{\sigma} & \s{2}{\alpha_3} & \none[\cdots] & \s{2}{\alpha_{d-2}}\\
\s{2}{\alpha_1}& \s{2}{\alpha_2} 
\end{ytableau} }
&
 even $d \geq 4$ 
 &
$(d-1)(d+2)(d+3)$\\
\;\;\\
  $ \hat{k}^{(d, 4)\mu \alpha_1 \nu \alpha_2 \rho \alpha_3 \sigma \alpha_4 \alpha_5 \ldots \alpha_{d-2}}  $ 
  &
\ytableausetup{centertableaux,boxsize=2.5em}
\scalebox{1}[.5]{
\begin{ytableau}
\s{2}{\mu} & \s{2}{\nu} & \s{2}{ \rho}  &\s{2}{\sigma} &  \s{2}{\alpha_4} & \s{2}{\alpha_5} & \none[\cdots] & \s{2}{\alpha_{d-2}}\\
\s{2}{\alpha_1}& \s{2}{\alpha_2} & \s{2}{\alpha_3} 
\end{ytableau} }
&
 even $d \geq 6$ 
 &
$\frac{5}{3}(d-3)(d+2)(d+1)$\\
\;\;\\
    \end{tabular}
    \end{ruledtabular}
\end{table*}

\subsection{GW propagations with Lorentz- and diffeomorphism-violating effects}

The equations of motion for GWs can be derived by varying the quadratic action $S \sim \int d^4 x \mathcal{L}$ with respect to $h_{\mu\nu}$ with Lagrangian density $\mathcal{L}$ given by (\ref{calL}), which yields
\bqn\lb{eom0}
\frac{1}{2}\eta_{\rho \sigma} \epsilon^{\mu\rho\alpha \kappa} \epsilon^{\nu\sigma \beta \lambda} \partial_\alpha \partial_\beta h_{\kappa \lambda} - \delta M^{\mu\nu \rho\sigma} h_{\rho \sigma}=0,
\eqn
where the tensor operators
\bqn
\delta M^{\mu\nu\rho \sigma} &=& - \frac{1}{4} \left(\hat{s}^{\mu\rho \nu \sigma} + \hat{s}^{\mu \sigma \nu \rho}\right) - \frac{1}{2} \hat{k}^{\mu \nu \rho \sigma} \nb\\
&&~~-\frac{1}{8} \left(\hat{q}^{\mu \rho \nu \sigma} + \hat{q}^{\nu \rho \mu \sigma} +\hat{q}^{\mu \sigma \nu \rho} + \hat{q}^{\nu \sigma \mu \rho}\right).\nb\\
\eqn
Here
\bqn
\hat{s}^{\mu\rho \nu \sigma}  = \sum_{d} \hat{s}^{(d)\mu\rho \nu \sigma}, \\
\hat{q}^{\mu\rho \nu \sigma}  = \sum_{d} \hat{q}^{(d)\mu\rho \nu \sigma}, \\
\hat{k}^{\mu\rho \nu \sigma}  = \sum_{d} \hat{k}^{(d)\mu\rho \nu \sigma}.
\eqn

In GR, the metric perturbation $h_{\mu \nu}$ contains only two degenerate, traceless, and transverse tensor modes. However, when Lorentz-violating and diffeomorphism-violating modifications are introduced, $h_{\mu \nu}$ may acquire additional degrees of freedom, depending on the specific nature of the violations. These include two scalar modes and two vector modes. As shown in \cite{Hou:2024xbv}, in Lorentz-violating linearized gravity, the extra scalar and vector modes of gravitational wave (GW) polarization can be directly induced by the two tensorial modes. {Here in this paper we restrict our study to the two transverse and traceless modes of GWs. Observationally, all the GW signals observed by the LIGO/Virgo/KAGRA collaboration exhibit compatibility with the two tensor polarization modes, showing no statistically significant signatures of additional polarization modes \cite{KAGRA:2021vkt}. Even if additional modes exist, their effects are expected to be  be subdominant compared to the two tensorial modes.} In this paper, assuming these additional modes are small, and following a similar treatment to that in \cite{Kostelecky:2016kfm}, we focus exclusively on the effects of Lorentz- and diffeomorphism-violating modifications on the two traceless and transverse tensor modes. {We note that this assumption imposes certain limitations, as our conclusions may not fully account for scenarios where such subdominant modes play a more substantial role. We left the analysis with the vector and scalar modes for future studies.}

With the above considerations, we restrict our analysis to the modes $h_{ij}$ that satisfy the conditions  
\bqn
\eta^{ij} h_{ij} = 0, \quad \partial^i h_{ij} = 0.
\eqn  
Under these constraints, the equations of motion for GWs given in Eq.~(\ref{eom0}) reduce to  
\bqn\lb{eom1}
(\partial_t^2 - \nabla^2) h^{ij} + 2 \delta M^{ij mn} h_{mn} = 0.
\eqn  
In the linearized gravity sector of SME, it is convenient to decompose the GWs into circular polarization modes. To study the evolution of $h_{ij}$, we expand it in terms of spatial Fourier harmonics as  
\bqn
h_{ij}(\tau, x^i) = \sum_{A={\rm R, L}} \int \frac{d^3 k}{(2\pi)^3} h_{\rm A}(\tau, k^i) e^{i k_i x^i} e_{ij}^{\rm A}(k^i),\nb\\
\eqn  
where $e_{ij}^{\rm A}$ are the circular polarization tensors, and R and L denote the right-handed and left-handed GW polarizations, respectively. The circular polarization tensors $e_{ij}^{\rm A}$ satisfy the relation 
\bqn
\epsilon^{ijk} n_j e_{kl}^A = i \rho_A e^{iA}_{~l},  
\eqn  
with $\rho_{\rm R} = 1$ and $\rho_{\rm L} = -1$.  
Using this decomposition, the equations of motion in Eq.~(\ref{eom1}) can be rewritten as  
\bqn
\ddot{h}_A + k^2 h_A + 2 \epsilon^A_{ij} \delta M^{ijmn} e^B_{mn} h_B = 0,
\eqn  
or equivalently, in matrix form:  
\begin{widetext}
\bqn
\left(
\begin{array}{cc}
    \partial_t^2 + k^2 + 2 e_{ij}^{R} \delta M^{ijmn} e^{R}_{mn} &  2 e_{ij}^{R} \delta M^{ijmn} e^{L}_{mn} \\
    2 e_{ij}^{L} \delta M^{ijmn} e^{R}_{mn}  &  \partial_t^2 + k^2 + 2 e_{ij}^{L} \delta M^{ijmn} e^{L}_{mn}
\end{array}
\right) 
\left(
\begin{array}{c}
   h_{R}  \\
   h_{L}
\end{array}
\right) = 0. \lb{eom_m}
\eqn
\end{widetext}  
Then, following methods developed for the study of Lorentz violation in the photon sector of the SME \cite{Kostelecky:2009zp} as well as GW propagations with diffeomorphism invariance, the modified dispersion relation of GWs with 4-momentum $k^{\mu} = (\omega, {\bf k})$ can be derived by requiring the determinant of the above $2\times 2$ matrix vanishes, which yields (see also in \cite{Kostelecky:2016kfm}) 
\footnote{It is worth noting here that the Lorentz- and diffeomorphism-violating operators introduced in (\ref{calL}) only affect the dispersion and the corresponding phase velocities of GWs. It is shown in \cite{Zhang:2024rel, Zhu:2022uoq, Zhu:2022dfq, Gao:2019liu} that the Lorentz-violating terms with mixed time and spatial derivatives can change the damping rates of GWs.}
\bqn
\omega = \Big(1- \zeta^0 \pm \left|\boldsymbol{\zeta}\right|\Big) |{\bf k}|,\lb{MD}
\eqn
where
\bqn \lb{zeta00}
\zeta^0 =  - \frac{1}{2 |{\bf k}|^2} \Big( e_{ij}^{R} \delta M^{ijmn} e^{R}_{mn} + e_{ij}^{L} \delta M^{ijmn} e^{L}_{mn}\Big),\nb\\
\eqn
and
\bqn \lb{zeta}
\left|\boldsymbol{\zeta}\right|^2 &=&  \frac{1}{4 |{\bf k}|^4} \Big[ (e_{ij}^{R} \delta M^{ijmn} e^{R}_{mn} - e_{ij}^{L} \delta M^{ijmn} e^{L}_{mn})^2 \nb\\
&&+ 4 (e_{ij}^{R} \delta M^{ijmn} e^{L}_{mn}) (e_{kl}^{L} \delta M^{klpq} e^{R}_{pq})\Big].
\eqn
And $\left|\boldsymbol{\zeta}\right|^2 \equiv (\zeta^1)^2 + (\zeta^2)^2 + (\zeta^3)^2$ with
\bqn
\zeta^1-i\zeta^2=\frac{1}{{\bf|k|}^2}(e_{ij}^{R}\delta M^{ijmn}e_{mn}^{L}),\\
\zeta^1+i\zeta^2=\frac{1}{{\bf|k|}^2}(e_{ij}^{L}\delta M^{ijmn}e_{mn}^{R}),
\eqn
\vspace{-2em}
\bqn
\zeta^3=\frac{1}{{2\bf|k|}^2}(e_{ij}^{R}\delta M^{ijmn}e_{mn}^{R}-e_{ij}^{L}\delta M^{ijmn}e_{mn}^{L}).
\eqn

The modified dispersion relation in the above leads to the phase velocities ($v=\omega/k$) of the GWs
\bqn
v_{\pm} = 1 - \zeta^0 \pm \left|\boldsymbol{\zeta}\right|.
\eqn
Here ``$\pm$" correspond to two modes propagating at different velocities.
Therefore, the two tensorial modes can be decompose into two modes propagate at different velocities, one is the fast mode (denoted by $h_f$ with velocity $v_+$) while another is the slow mode (denote by $h_s$ with velocity $v_-$). 

{Now we need to connect ($h_f,h_s$) with the circular polarization modes ($h_{L},h_{R}$). For this purpose, one can substitute the dispersion relation in Eq.~(\ref{MD}) to the equation of motion (\ref{eom_m}) for fast and slow modes respectively. Note that in the substitution, we use the relation $\partial_t^2 = - \omega^2$. For fast mode, by using the experssions of $\zeta_0$ and ${\boldsymbol \zeta}$ in Eqs.~(\ref{zeta00}) and ~(\ref{zeta}), one has
\bqn
2|{\bf k}|^2 
\left(
\begin{array}{cc}
    \zeta^3 -\left|\boldsymbol{\zeta}\right| &  \zeta^1-i \zeta^2 \\
    \zeta^1+i\zeta^2 & -\zeta^3-\left|\boldsymbol{\zeta}\right|
\end{array}
\right) 
\left(
\begin{array}{c}
   h_{R}  \\
   h_{L}
\end{array}
\right) = 0.
\eqn
This equation admits a set of solution in the form of
\bqn \lb{fast}
\left(
\begin{array}{c}
    h_R\\
    h_L
\end{array}
\right)_{\rm fast} =
\left(
\begin{array}{c}
   e^{-i \varphi/2} \cos\frac{\vartheta}{2} \\
   e^{ i \varphi/2} \sin\frac{\vartheta}{2}
\end{array}
\right),
\eqn
where} the angles $\varphi$ and $\vartheta$ are defined as
\bqn
\sin\vartheta &=& \frac{\sqrt{(\zeta^1)^2+(\zeta^2)^2}}{\left|\bm{\zeta}\right|
}, \\
\cos\vartheta &=& \frac{\zeta^3}{\left|\boldsymbol{\zeta}\right|}, \\
e^{\pm i\varphi} &=& \frac{\zeta^1\pm \zeta^2}{\sqrt{(\zeta^1)^2+(\zeta^2)^2}}. 
\eqn
{Similarly, for slow mode, one has
\bqn \lb{slow}
\left(
\begin{array}{c}
    h_R\\
    h_L
\end{array}
\right)_{\rm slow} =
\left(
\begin{array}{c}
   -e^{-i \varphi/2} \sin\frac{\vartheta}{2} \\
   e^{ i \varphi/2} \cos\frac{\vartheta}{2}
\end{array}
\right).
\eqn
Therefore, Eq.~(\ref{fast}) and (\ref{slow}) can be written as a unique equation as }
\bqn
{
\left(
\begin{array}{ccc}
   h_R  \\
h_L
\end{array}
\right)  = 
\left(
\begin{array}{ccc}
       e^{-i\varphi/2}\cos\frac{\vartheta}{2}
  &  - e^{-i\varphi/2}\sin\frac{\vartheta}{2}  \\
e^{i\varphi/2}\sin\frac{\vartheta}{2}   &   e^{i\varphi/2}\cos\frac{\vartheta}{2} 
\end{array}
\right) 
\left(
\begin{array}{ccc}
   h_f  \\
h_s
\end{array}
\right).}\nb\\ 
\eqn
{It is also convenient to express $(h_f, h_s)$ in terms of $(h_R , h_L)$ as}
\bqn\label{fast_slow_modes}
\left(
\begin{array}{ccc}
   h_f  \\
h_s
\end{array}
\right)  = 
\left(
\begin{array}{ccc}
       e^{i\varphi/2}\cos\frac{\vartheta}{2}
  &  e^{-i\varphi/2}\sin\frac{\vartheta}{2}  \\
-e^{i\varphi/2}\sin\frac{\vartheta}{2}   &   e^{-i\varphi/2}\cos\frac{\vartheta}{2} 
\end{array}
\right) 
\left(
\begin{array}{ccc}
   h_{R} \\
h_{L}
\end{array}
\right).\nb\\ 
\eqn
The circular polarization modes $h_{R}$ and $h_{L}$ relate to the modes $h_+$ and $h_\times$ via
\bqn
h_+ = \frac{h_{L}+h_{R}}{\sqrt{2}}, \\
h_\times = \frac{h_{L}- h_{R}}{\sqrt{2 }i}.
\eqn
{Then we can write $(h_f, h_s)$ in terms of $h_+$ and $h_\times$ as
\begin{widetext}
\begin{equation}
\begin{pmatrix} 
h_f \\ 
h_s 
\end{pmatrix} =
\frac{1}{\sqrt{2}}
\begin{pmatrix}
 -e^{i\varphi/2}\sin\frac{\vartheta}{2} + e^{-i\varphi/2}\cos\frac{\vartheta}{2}  & ie^{i\varphi/2}\sin\frac{\vartheta}{2}+ ie^{-i\varphi/2}\cos\frac{\vartheta}{2}  \\
 e^{i\varphi/2}\cos\frac{\vartheta}{2} + e^{-i\varphi/2}\sin\frac{\vartheta}{2}  &  -ie^{i\varphi/2}\cos\frac{\vartheta}{2} +ie^{-i\varphi/2}\sin\frac{\vartheta}{2} )
\end{pmatrix}
\begin{pmatrix} 
h_+ \\ 
h_\times 
\end{pmatrix}.
\end{equation}
And inversely, 
\begin{equation}
{
\begin{pmatrix}
h_+ \\
h_\times
\end{pmatrix}
=
\frac{1}{\sqrt{2}}
\begin{pmatrix}
 e^{i\varphi/2}\cos\frac{\vartheta}{2} - e^{-i\varphi/2}\sin\frac{\vartheta}{2} & e^{-i\varphi/2}\cos\frac{\vartheta}{2} + e^{i\varphi/2}\sin\frac{\vartheta}{2} \\
 -ie^{-i\varphi/2}\sin\frac{\vartheta}{2} - ie^{i\varphi/2}\cos\frac{\vartheta}{2} & -ie^{i\varphi/2}\sin\frac{\vartheta}{2} + ie^{-i\varphi/2}\cos\frac{\vartheta}{2}
\end{pmatrix}
\begin{pmatrix}
h_f \\
h_s
\end{pmatrix}.
}
\end{equation}
\end{widetext}}

In the modified dispersion relation (\ref{MD}), the coefficient $\zeta^0$ and ${\bf \zeta}$ are functions of the frequency $\omega$ and wave vector ${\bf k}$ \cite{Niu:2022yhr}. Considering it's also direction-dependent and to describe its effects on the propagation of GWs, it is convenient to expand its coefficients in terms of spin-weighted spherical harmonics $\;_sY_{jm}({\bf \hat n})$ as
\bqn
\zeta^0 &=& \sum_{d, jm} \omega^{d-4} Y_{jm}({\bf \hat n}) k_{( I)jm}^{(d)}, \\
\zeta^1\mp i\zeta^2 &=& \sum_{d, jm} \omega^{d-4} \;_{\pm4}Y_{jm}({\bf \hat n}) \Big[k_{(E)jm}^{(d)}\pm k_{(B)jm}^{(d)}\Big], \nb\\
&&\;\; \\
\zeta^3 &=& \sum_{d, jm} \omega^{d-4} Y_{jm}({\bf \hat n}) k_{(V)jm}^{(d)},
\eqn
where ${\bf n} = - {\bf k}$ is the direction of the source, $Y_{jm}({\bf \hat n})=\;_0Y_{jm}({\bf \hat n})$ is the scalar spherical harmonics function, and $|s| \leq j \leq d-2$. The indice $m$ takes $-j, \cdots, j$. The spherical coefficients for Lorentz and diffeomorphism violations $k_{(I)jm}^{(d)}$, $k_{(E)jm}^{(d)}$, $k_{(B)jm}^{(d)}$, and $k_{(V)jm}^{(d)}$ are linear combinations of the tensor coefficients in Eqs. (\ref{sd}, \ref{qd}, \ref{kd}), which obey the relation $k_{jm}^{(d)*}=(-1)^m k_{j -m}^{(d)}$. The expansions of the coefficient $\zeta^0$, $\zeta^1 \pm i\zeta^2$, and $\zeta^3$ are also a combination of operators at multiple mass dimensions, but they have different expansion properties. The mass dimension $d$ of the expansion coefficients can take even numbers of $d\geq 2$ for the expansion of $\zeta^0$, odd numbers of $d \geq 3$ for $\zeta^3$, and even number of $d \geq 6$ for $\zeta^1 \pm i\zeta^2$.

For convenience, the frequency and direction dependence can be separated, and we introduce several energy-independent coefficients as
\bqn
&&\zeta^0_{(d)} ({\bf n}) = \sum_{jm}Y_{jm}({\bf \hat n}) k_{(I)jm}^{(d)}, \lb{zeta0} \\
&&\zeta^1_{(d)} ({\bf n}) \mp i\zeta^2_{(d)}({\bf n})  \nb\\
&&~~~~~~~ = \sum_{jm}\;_{_{\pm4}}Y_{jm}({\bf \hat n}) \Big[k_{(E)jm}^{(d)} \pm ik_{(B)jm}^{(d)}\Big], \lb{zeta12}\\
&&\zeta^3_{(d)} ({\bf n}) = \sum_{jm}Y_{jm}({\bf \hat n}) k_{(V)jm}^{(d)}.\lb{zeta3}
\eqn
Then the phase velocity of the GWs can be rewritten as
\bqn
v_{\pm}  &=& 1 - \omega^{d-4}\Big[ \zeta^0_{(d)}({\bf n}) \nb\\ && \mp \sqrt{(\zeta^1_{(d)}{(\bf n)}^2+(\zeta^2_{(d)}{(\bf n)}^2+(\zeta^3_{(d)}{(\bf n)}^2}\Big],  \lb{zeta0}
\eqn
for a specific mass dimension $d$. The new effects arising from the Lorentz and diffeomorphism violations in the linearized gravity of SME are fully characterized by the coefficients, $k^{(d)}_{(I)jm}$, $k^{(d)}_{(V)jm}$, and $k^{(E)}_{(E)jm} \pm i k^{(d)}_{(B)jm}$. These coefficients determine the speed of the GWs and all lead to frequency-dependent dispersions. Specifically, the coefficients $k^{(d)}_{(V)jm}$ and $k^{(E)}_{(I)jm} \pm i k^{(d)}_{(B)jm}$ lead to different velocities of two independent tensorial modes of GWs, a fast mode $h_f$ and a slow mode $h_s$, as we mentioned in Eq.~(\ref{fast_slow_modes}). Therefore, the arrival times of the fast and slow modes could be different. This is also called the velocity birefringence of the GWs. The coefficients $k^{(d)}_{(I)jm}$ induce the non-birefringent dispersion of GWs. For this case (except $d=4$ case), the two independent tensorial modes have the frequency-dependent velocity in the same form. For $d=4$ case, the velocity is independent of the frequency of the GWs. In addition, all the coefficients, $k^{(d)}_{(I)jm}$, $k^{(d)}_{(V)jm}$, and $k^{(d)}_{(E)jm} \pm i k^{(d)}_{(B)jm}$ are also direction-dependent if $j \neq 0$ and induce the anisotropic phase effects on the propagation of the GWs. 

In summary, all the coefficients can provide frequency and direction-dependent phase modifications to the GWs. In the following, we are going to study the phase modifications due to these Lorentz- and diffeomorphism-violating coefficients in detail.

\section{Phase Modifications to the Waveform of GWs} \lb{sec3}
\renewcommand{\theequation}{3.\arabic{equation}} \setcounter{equation}{0}

{In this section, we consider the propagation of GWs with Lorentz- and   diffeomorphism-violating effects in  a Friedman-Robertson-Walker (FRW) background, whose metric is given by,
\begin{equation}
    ds^2 = -dt^2 + a^2(t) d\mathbf{r}^2,
\end{equation}
where $t$ is the cosmic time, $a(t)$ is the scale factor governing the expansion of the universe.} Now we consider a graviton emitted radially at $r=r_e$ and received at $r=0$, we have
\bqn
\frac{dr}{dt} = - \frac{1}{a} \Big[1 - \zeta^0\pm \sqrt{(\zeta^1)^2 + (\zeta^2)^2 + (\zeta^3)^2} \Big].
\eqn
Integrating this equation from the emission time (when $r=r_e$) to arrival time (when $r=0$), one obtains
\bqn
r_e&=&  \int_{t_e}^{t_0} \frac{dt}{a(t)}- \omega^{d-4} \Big[\zeta^0_{(d)}({\bf n})\nb\\&& \mp \sqrt{(\zeta^1_{(d)}{(\bf n))}^2+(\zeta^2_{(d)}{(\bf n))}^2+(\zeta^3_{(d)}{(\bf n))}^2}\Big] \int_{t_e}^{t_0} \frac{dt}{a^{d-3}}.\nb\\
\eqn
Considering gravitons emitted at two different times $t_e$ and $t_e'$, with wave numbers $k$ and $k'$, and received at corresponding arrival times $t_0$ and $t_0'$ ($r_e$ is the same for both). Assuming $\Delta \equiv t_e - t'_e \leq a/\dot{a} $,  then, the difference in their arrival times is given by 
\bqn
&&\Delta t_0 = (1+z) \Delta t_e + \Big(\omega^{d-4}_e - \omega'^{d-4}_e\Big) \Big[\zeta^0_{(d)}({\bf n})\nb\\
&& \mp \sqrt{(\zeta^1_{(d)}{(\bf n))}^2+(\zeta^2_{(d)}{(\bf n))}^2+(\zeta^3_{(d)}{(\bf n))}^2}\Big] \int_{t_e}^{t_0} \frac{dt}{a^{d-3}},\nb\\
\lb{time}
\eqn
where $z = 1/a(t_e) -1$ is the cosmological redshift. 

Let us focus on the GW signal generated by non-spinning, quasi-circular inspiral in the post-Newtonian approximation. Relative to the GW in GR, the Lorentz- and diffeomorphism-violating effects  modify the phase of GWs. {The phase corrections to the GR-based waveform due to Lorentz- and diffeomorphism-violating effects can be computed using the stationary phase approximation (SPA) during the inspiral phase of the binary system \cite{Zhao:2019xmm, Mirshekari:2011yq}. As demonstrated in \cite{Ezquiaga:2022nak}, the waveforms modified by propagation effects and derived using the SPA agree with those derived using the WKB approximation. In WKB approximation, the corrections to the GR-based waveform are only due to the propagation effect, and thus it is in principle independent of the GW emission mechanism or radiated stages of the binary system \cite{Ezquiaga:2022nak}. This implies that one can extend the modified waveforms obtained using the SPA \cite{Zhao:2019xmm, Mirshekari:2011yq} to the entire signal including the inspiral, merger, and ringdown phases of a coalescing binary system. }

{In the SPA, the Fourier transform of $h_A(t)$ can be obtained analytically, which is given by \cite{Maggiore:2007ulw}
\bqn
\tilde{h}_A(f) = \frac{{\cal A}_A(f)}{\sqrt{df/dt}}e^{i \Psi(f)},
\eqn
where $f$ is the GW frequency at the detector, ${\cal A}_A(f)$ is the amplitude, and $\Psi$ is the phase of GWs.} In \cite{Will:1997bb, Mirshekari:2011yq}, it was proved that the difference of arrival times in (\ref{time}) induces the modification to the phase of GWs $\Psi$ in the following form,
{
\bqn
\Psi(f) =\Psi^{\rm GR} (f) + \delta \Psi(f),
\eqn
where $\Psi^{\rm GR} (f)$ is the phase predicted in GR and $\delta \Psi(f)$ is the phase corrections due to the Lorentz- and diffeomorphism-violating effects. According to the time difference in Eq.~(\ref{time}), the phase corrections can be divided into two parts, 
\bqn
\delta \Psi(f) =\mp \delta \Psi_1(f, {\bf n}) + \delta \Psi_2(f, {\bf n}),
\eqn
where $\delta \Psi_1(f, {\bf n})$ corresponds to the velocity birefringence effect and $\delta \Psi_2(f, {\bf n})$ represents the non-birefringent effects. $\delta \Psi_1$ is induced by the Lorentz- and diffeomorphism-violating coefficients $k^{(d)}_{(V)jm}$ and $k^{(E)}_{(I)jm} \pm i k^{(d)}_{(B)jm}$, while $\delta \Psi_2$ is induced by $k^{(d)}_{(I)jm}$.}

For $d\neq 3$, where $\mp$ correspond to fast and slow modes, respectively, and
\bqn
&&\delta \Psi_1(f, {\bf n})  = \frac{2^{d-3}}{d-3} \frac{u^{d-3}}{\mathcal{M}^{d-3}} \nb\\  
&&\sqrt{(\zeta^1_{(d)}{(\bf n))}^2+(\zeta^2_{(d)}{(\bf n))}^2+(\zeta^3_{(d)}{(\bf n))}^2} \int_{t_e}^{t_0} \frac{dt}{a^{d-3}},\nb\\
\eqn
\vspace{-2.5em} %
\bqn
 &&\delta \Psi_2(f, {\bf n})  = \frac{2^{d-3}}{d-3} \frac{u^{d-3}}{\mathcal{M}^{d-3}}  \zeta^0_{(d)}({\bf n})  \int_{t_e}^{t_0} \frac{dt}{a^{d-3}}, \lb{Phi}
\eqn
where $u=\pi \mathcal{M} f$ with $f=\omega/2 \pi$ being the frequency of the GWs, $\mathcal{M} = (1+z) \mathcal{M}_c$ is the measured chirp mass, and $\mathcal{M}_c \equiv (m_1 m_2)^{3/5}/(m_1+m_2)^{1/5}$  is the chirp mass of the binary system with component masses $m_1$ and $m_2$. When $d=3$, the phase corrections $\delta \Psi_1$ and $\delta \Psi_2$ are given by
\bqn
\delta \Psi_1 &=& \sqrt{(\zeta^1_{(d)}{(\bf n))}^2+(\zeta^2_{(d)}{(\bf n))}^2+(\zeta^3_{(d)}{(\bf n))}^2} \nb\\
&&~~~~~~ \times \ln{u}  \int_{t_e}^{t_0} dt, \\
\delta \Psi_2 &=&  \zeta^0_{(d)}({\bf n}) \ln{u}  \int_{t_e}^{t_0} dt.
\eqn
In this paper, we adopt the above modified waveforms for later analysis with the open data of compact binary merging events detected by the LIGO-Virgo-KAGRA collaboration.

Let us turn to derive the modified waveform of GW with Lorentz- and diffeomorphism-violating effects in the linearized gravity sector of the SME. For this purpose, we closely follow the derivation presented in \cite{Qiao:2019wsh, Zhao:2019xmm, Mewes:2019dhj}.Then the fast and slow modes $(h_f , h_s)$ with phase corrections can expressed as
\bqn
h_f = h_f^{\rm GR}e^{(\delta \Psi_2 + \delta \Psi_1)} , \\
h_s = h_f^{\rm GR}e^{(\delta \Psi_2 - \delta \Psi_1)}.
\eqn
By using the relation in (\ref{fast_slow_modes}), one can transform the above results to the circular polarization modes
\bqn\lb{h_R}
&&h_{R} = e^{i\delta \Psi_2} \Big[(\cos\delta \Psi_1 - i\cos\vartheta \sin\delta \Psi_1)h_{R}^{\text{GR}} \nb\\ 
&&~~~~~~~~- i\sin\vartheta e^{-i\varphi}\sin\delta \Psi_1 h_{L}^{\text{GR}} \Big] ,\\
&&h_{L} = e^{i\delta \Psi_2} \Big[(\cos\delta \Psi_1 + i\cos\vartheta \sin\delta \Psi_1)h_{L}^{\text{GR}} \nb\\ 
&&~~~~~~~~- i\sin\vartheta e^{i\varphi}\sin\delta \Psi_1 h_{R}^{\text{GR}} \Big] .\lb{h_L}
\eqn
The corresponding waveforms for $h_+$ and $h_\times$ are
\bqn
&&h_+ = e^{i\delta \Psi_2} \Big[(\cos\delta \Psi_1 - i\cos\varphi\sin\vartheta \sin\delta \Psi_1)h_+^{\text{GR}} \nb\\ 
&&~~~~~~~ -(\cos\vartheta+i\sin\vartheta\sin\varphi) \sin\delta\Psi_1h_\times^{\text{GR}} \Big], \lb{hplus}\\
&& h_\times = e^{i\delta \Psi_2} \Big[(\cos\delta \Psi_1 + i\cos\varphi\sin\vartheta \sin\delta \Psi_1)h_\times^{\text{GR}} \nb\\ 
&&~~~~~~~ +(\cos\vartheta-i\sin\vartheta\sin\varphi) \sin\delta\Psi_1h_+^{\text{GR}} \Big]\lb{hcross}.
\eqn
{Here we would like to add two remarks about the above modified waveforms and their tests with GW signals detected by LIGO-Virgo-KAGRA detectors. Firstly, the modified waveform presented in the above represents the most general waveform with Lorentz- and diffeomorphism-violating effects in the linearized gravity.  Here we note that the modified waveforms with Lorentz-violating but diffeomorphism invariant coefficients for $d>4$ has been considered in refs. \cite{Gong:2023ffb, Kostelecky:2016kfm, Haegel:2022ymk, Niu:2022yhr, ONeal-Ault:2021uwu}. The waveforms with phase corrections presented above generalize those in Refs.~\cite{Gong:2023ffb, Kostelecky:2016kfm, Haegel:2022ymk, Niu:2022yhr, ONeal-Ault:2021uwu} to the Lorentz- and diffeomorphism-violating case, valid for $d \geq 2$. Secondly, the constraints on the Lorentz-violating coefficients with mass dimensions $d=5$ and $d=6$ have been studied by comparing the above modified waveforms with GW signals detected by LIGO-Virgo-KAGRA detectors have been performed in refs.~\cite{Gong:2023ffb, Haegel:2022ymk, Niu:2022yhr}. And the main purpose of the rest part of this paper is to use the above modified waveforms to derive the constraints on the Lorentz- and diffeomorphism-violating coefficients for $d=2$ and $d=3$ cases.}

Then we would like to consider several special limits of the above general waveform. Considering that the operators with the lowest mass dimension are expected to have the
dominant Lorentz- and diffeomorphism-violating effects on the propagation of GWs, in the following we only discuss specific cases with relative lower mass dimensions, for example, $d=2, 3, 4, 5, 6$. Note that ref.~\cite{Gong:2021jgg} explores isotropic Lorentz-violating effects on GWs, which correspond to operators with mass dimensions $d=7$ and $d=8$. 

\subsection{Non-birefringent dispersion by the even $d$ coefficients $k_{(I)jm}^{(d)}$}

The coefficients $k_{(I)jm}^{(d)}$ for even $d \geq 2$ induce the non-birefringent dispersion of GWs,
\bqn
\omega = \left(1 - \omega^{d-4} \sum_{jm} Y_{jm}({\bf \hat n})k_{(I)jm}^{(d)} \right) |{\bf k}|.
\eqn
This leads to the non-birefringent phase velocity of GWs,
\bqn
v = 1 - \omega^{d-4} \sum_{jm} Y_{jm}({\bf \hat n})k_{(I)jm}^{(d)}.
\eqn
This phase velocity is direction-independent for $j=0$ but anisotropic for $j \neq 0$. For the effect induced by $k_{(I)jm}^{(d)}$, it is obvious to see that the phase modification $\delta \Psi_1=0$. In the following, we consider the phase corrections for mass dimensions $d=2$, $d=4$, and $d=6$, respectively.

\subsubsection{d=2}

For mass dimension $d=2$ case, since $0 \leq j \leq d-2$, $j$ and $m$ have to take $j=0=m$. For this case, $k_{(I)jm}^{(2)}$ only has one component $k_{(I)00}^{(2)}$, which is obvious direction independent. The phase correction for this case is given by 
\bqn
\delta \Psi_2 = - \frac{(\pi f)^{-1}}{4 \sqrt{\pi} } k_{(I)00}^{(2)} \int_{t_e}^{t_0} a dt.\lb{phase_d2}
\eqn
As we mentioned, the $d=2$ case can only be induced by the diffeomorphism violations of the linearized gravity described by the Lagrangian (\ref{calL}). 

\subsubsection{d=4}

For the $d=4$ case, the phase velocity is independent of the frequency of the GWs, so they do not give any observable dephasing effects and modify the speed of GWs in a frequency-independent way. This effect can be constrained by comparison with the arrival time of the photons from the associated electromagnetic counterpart. For the binary neutron star merger GW170817 and its associated electromagnetic counterpart GRB170817A \cite{LIGOScientific:2017zic}, the almost coincident observation of the electromagnetic wave and the GW place an exquisite bound on
\bqn
-7 \times 10^{-17}< \sum_{jm} Y_{jm}({\bf \hat n})k_{(I)jm}^{(4)} < 3 \times 10^{-15}.
\eqn
Since this case does not lead to any dephasing effects, we are not going to include this case in our later analysis with GW signals in GWTC-3. In addition, it is shown in \cite{Hou:2024xbv} that with the effects of the anisotropic coefficients $k_{(I)jm}^{(4)}$, the extra polarizations of GWs can be directly generated by the two tensorial modes under certain conditions.

\subsubsection{d=6}

For the mass dimension $d=6$ case, the index $j$ can take 0, 1, 2, 3, 4, and the index $m$ runs from $-j$ to $j$. Note that each of $k_{(I)jm}^{(6)}$ are complex function which satisfies $k_{(I)jm}^{(6)*}=(-1)^{m} k_{(I)j-m}^{(6)}$. Thus the number of independent components for coefficients $k_{(I)jm}^{(6)}$ are $(d-1)^2=25$. The phase correction in the modified waveform for this case is given by 
\bqn
\delta \Psi_2 = \frac{8}{3} (\pi f)^3 \left(\sum_{jm} Y_{jm}({\bf \hat n})k_{(I)jm}^{(6)} \right) \int_{t_e}^{t_0} \frac{dt}{a^3}.
\eqn
The coefficients $k_{(I)jm}^{(6)}$ can be constrained by comparing the modified waveform with the GW strain data from the GW detectors, see ref. \cite{Qiao:2022mln} for a review. The analysis with the isotropic effect of Lorentz violation which corresponds to $j=0$ in the linearized gravity has been carried out through full Bayesian parameter estimations on the GW events observed by the LIGO/Virgo/KAGRA detectors in a series of papers \cite{Zhu:2023rrx, LIGOScientific:2019fpa, LIGOScientific:2020tif, LIGOScientific:2021sio, Zhu:2022uoq}, while the anisotropic case has been considered in \cite{Gong:2023ffb}.

\subsection{Birefringent dispersion by the odd $d$ coefficients $k_{(V)jm}^{(d)}$}

The coefficients $k_{(V)jm}^{(d)}$ for odd $d \geq 3$ produces frequency-dependent dispersion and birefringence effects in GWs, i.e.,
\bqn
\omega = \left(1 \pm \omega^{d-4} \sum_{jm} Y_{jm}({\bf \hat n})k_{(V)jm}^{(d)} \right) |{\bf k}|.
\eqn
Note that here $\mp$ corresponds to the fast and slow modes respectively. This leads to the birefringent phase velocity of GWs,
\bqn
v_{\pm} = 1 \pm \omega^{d-4} \sum_{jm} Y_{jm}({\bf \hat n})k_{(V)jm}^{(d)}.
\eqn
Similarly, this velocities are direction-independent for $j=0$ but anisotropic for $j \neq 0$. In this case, the fast and slow modes are circularly polarized, and we have $\vartheta=0,\pi$. For the effects induced by $k_{(V)jm}^{(d)}$, it is obvious that the phase corrections $\delta \Psi_2=0$.

\subsubsection{d=3}

For mass dimension $d=3$ case, the index $j$ can take 0 and 1, and index $m$ runs from $-j$ to $j$. Considering $k_{(V)jm}^{(5)*}=(-1)^{m} k_{(V)j-m}^{(5)}$, it is known that there are only 4 independent components for the coefficients $k_{(V)jm}^{(3)}$, they are $k_{(V)00}^{(3)}$, $k_{(V)10}^{(3)}$, ${\rm Re}k_{(V)11}^{(3)}$, and ${\rm Im}k_{(V)11}^{(3)}$, in which $k_{(V)11}^{(3)}$ is a complex function so it contains two independent components. The phase correction for this case is given by 
\bqn
\delta \Psi_1 = \ln u  \left(\sum_{jm} Y_{jm}({\bf \hat n})k_{(V)jm}^{(3)} \right) \int_{t_e}^{t_0} dt.
\eqn
Similarly, one can constrain the coefficients $k_{(V)jm}^{(3)}$ by comparing the modified waveform with the GW strain data from the GW detectors. The analysis with the isotropic effect for $d=3$ case (which corresponds to $j=0$ case) has been performed in refs.~\cite{Zhu:2023rrx, Zhu:2022uoq, Wu:2021ndf}. 

\subsubsection{d=5}

For mass dimension $d=5$ case, the index $j$ can take 0, 1, 2, 3, and index $m$ runs from $-j$ to $j$. Here the coefficients $k_{(V)jm}^{(5)}$ have $(d-1)^2=16$ components in total. The phase corrections for this case are given by
\bqn
\delta \Psi_1 = 2(\pi f)^2 \left(\sum_{jm} Y_{jm}({\bf \hat n})k_{(V)jm}^{(3)} \right) \int_{t_e}^{t_0} a^{-2} dt.
\eqn
The constraints on the isotropic effect for $d=5$ case have been performed in a series of papers, see refs.~\cite{Zhu:2023rrx, Zhu:2022uoq, Wang:2020cub, Zhao:2022pun, Wang:2021gqm}, while the the anisotropic effects has been constrained in \cite{Niu:2022yhr, Haegel:2022ymk, Shao:2020shv, Wang:2021ctl}.

\subsection{Birefringent dispersion by the even $d$ coefficients $k_{(E)jm}^{(d)}$ and $k_{(B)jm}^{(d)}$}

The coefficients $k_{(E)jm}^{(d)}$ and $k_{(B)jm}^{(d)}$ for even $d \geq 6$ produces frequency-dependent dispersion and birefringence effects in GWs, i.e.,
\bqn
\omega = \left(1 \pm \frac{1}{2}\omega^{d-4} \sqrt{(\zeta^1_{(d)} ({\bf n}))^2 + (\zeta^2_{(d)}({\bf n}))^2 }\right) |{\bf k}|.
\eqn
For this case, we have $\vartheta=\pi/2$ and $\delta \Psi_2=0$. 

For mass dimension $d=6$, the index $j=4$ and index $m$ runs from $0$ to $4$. In this case, the  coefficients $k_{(E)jm}^{(d)}$ and $k_{(B)jm}^{(d)}$ have 18 components in total. The phase corrections for this case is given by
\bqn
\delta \Psi_1 = \frac{8}{3}(\pi f)^3 \sqrt{(\zeta^1_{(6)} ({\bf n}))^2 + (\zeta^2_{(6)}({\bf n}))^2 }\int_{t_e}^{t_0} a^{-3} dt.\nb\\
\eqn
The constraints on the coefficients $k_{(E)jm}^{(d)}$ and $k_{(B)jm}^{(d)}$ for $d=6$ case have been performed in \cite{Niu:2022yhr, Shao:2020shv, Wang:2021ctl}.

\section {Bayesian inference and parameter estimation}
\lb{sec4}
\renewcommand{\theequation}{4.\arabic{equation}} \setcounter{equation}{0}

In this section, we present a brief introduction to the Bayesian inference used to constrain the coefficients of Lorentz and diffeomorphism violation in the linearized gravity in the framework of the SME. Bayesian inference plays a pivotal role in modern astronomy, enabling the extraction of physical parameters from observational data.  Given GW data $d_i$, we compare it with the predicted GW strain incorporating Lorentz- and diffeomorphism-violating effects to infer the posterior distribution of parameters $\vec{\theta}$ that characterize the waveform model. According to Bayes' theorem, the posterior distribution is expressed as:
\bqn
 P({\vec \theta}|d,H)=\frac{P(d| \vec{\theta},H) P(\vec{\theta}| H)}{P(d|H)},
 \eqn
where $ P({\vec \theta}|d, H)$ represents the posterior probability distributions of the model parameters ${\vec{\theta}}$, and $H$ is the waveform model. $P(\vec{\theta}| H)$ is the prior distribution for parameters $\vec{\theta}$, $P(d| \vec{\theta}, H)$ is the likelihood function for a given set of model parameters and $P(d|H)$ is the normalization factor, commonly referred as the ``evidence":
 \bqn
 P(d|H) \equiv \int d \vec{\theta} P(d| \vec{\theta}, H) P(\vec{\theta}| H).
 \eqn

 In most cases, GW signals are weak and embedded within noise, making matched filtering a crucial method for signal extraction. Assuming Gaussian and stationary noise \cite{Cutler:1994ys, Romano:2016dpx, Thrane:2018qnx}, the likelihood function for matched filtering is given by:
\bqn
P(\boldsymbol{d}|{\boldsymbol{\theta}}, H) \propto \prod_{i=1}^{n} e^{-\frac{1}{2}\langle \boldsymbol{d_i}-\boldsymbol{h({\theta})}|\boldsymbol{d_i}-\boldsymbol{h({\theta})}\rangle},
\eqn
where $\boldsymbol{h({\theta})}$ is the GW strain predicted by the waveform model $H$, and $i$ indexes the individual GW detectors. The noise-weighted inner product $\langle A|B \rangle$ is defined as:
\bqn
\langle A|B \rangle = 4\; {\rm Re} \left[\int_0^\infty \frac{A(f) B(f)^*}{S(f)} df\right],
\eqn
where $^*$ denotes the complex conjugate, and  {$A(f)$ represents the measured GW strain signal at the LIGO/Virgo/KAGRA detectors, and $B(f)$ represents the theoretical GW predicted by the waveform model.} $S(f)$ is the power spectral density (PSD) of the detector. To ensure stable and reliable parameter estimation, we use the PSD data provided in the LVK posterior sample, which is more robust compared to PSDs derived from strain data using Welch averaging \cite{LIGOScientific:2018mvr, Cornish:2014kda, Littenberg:2014oda}.

We consider the cases of Lorentz- and diffeomorphism-violating waveforms in (\ref{hplus}) and (\ref{hcross}) with different mass dimension $d$ separately. It is mentioned in the previous section that the cases with mass dimensions $d=5$ and $d=6$ have been explored in refs. \cite{Gong:2023ffb, Niu:2022yhr, Haegel:2022ymk, Wang:2021ctl, Shao:2020shv}, which will not be considered here. For this reason, we explore the cases with $d=2$ and $d=3$ cases in this paper. To perform the parameter estimation on the modified waveforms (\ref{hplus}) and (\ref{hcross}) with the Lorentz- and diffeomorphism-violating effects, we employ the python package BILBY \cite{Romero-Shaw:2020owr, Ashton:2018jfp}. We perform Bayesian inference on GW data from the 88 compact binary coalescence events in GWTC-3, which include binary neutron stars like GW170817, neutron star–black hole binaries, and binary black holes. 
Two events, GW200308\_173609 and GW200322\_091133, are excluded in our analysis due to the possible uncertainties of their inferred source properties \cite{KAGRA:2021vkt}. It is also shown in \cite{Morras:2022ysx} from a new analysis that these two events could be generated by Gaussian noise fluctuations.
We use the \texttt{IMRPhenomXPHM} template \cite{Garcia-Quiros:2020qpx, Pratten:2020ceb, Pratten:2020fqn} for the GR waveform $h^{\rm GR}_{+, \; \times}(f)$ except for the binary neutron star event GW170817, and use \texttt{IMRPhenomPv2\_NRTidal} for GW170817.

Considering that the spherical expansion coefficient formulae in Eqs.~(\ref{zeta0}, \ref{zeta12}, \ref{zeta3}) provide a general solution for different events within the same coordinate system, we can combine the posterior distributions of individual events as:
\bqn
P(\vec{\theta}|\{d_i\}, H) \propto \prod_{i=1}^{N} P(\vec{\theta}| d_i, H), \lb{combine}
\eqn
where $d_i$ represents the data from the $i$-th GW event, and $N$ is the total number of selected GW events.

\section {Results}
\renewcommand{\theequation}{5.\arabic{equation}} \setcounter{equation}{0}
\lb{sec5}

In this section, we present the results of the constraints on the Lorentz- and diffeomorphism-violating coefficients for mass {dimensions} $d=2$ and $d=3$ cases. In the following, we present the results for $d=2$ and $d=3$ separately. 

\subsection{$d=2$}

\begin{figure}
\centering
\includegraphics[width=8cm]{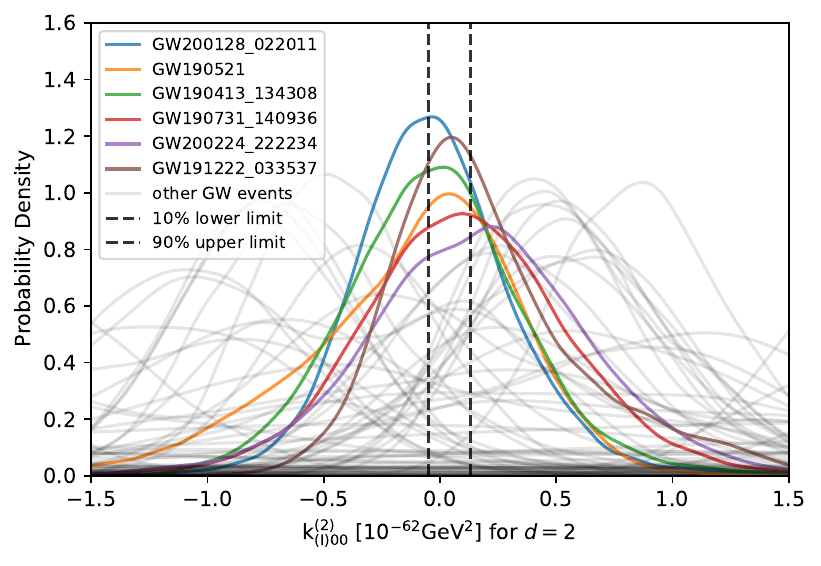}
\caption{The posterior distributions for $ k_{(I)00}^{(2)}$ from selected GW events in the LIGO-Virgo-KAGRA catalog GWTC-3. The legend indicates the events that give the six tightest constraints. The vertical dash line denotes the 90\% interval for $k_{(I)00}^{(2)}$ from combined results. Note that we have excluded a few events in our analysis (as shown in the list presented in Table.~\ref{excluded_list}). 
\label{kI00}}
\end{figure}

\begin{figure}
\centering
\includegraphics[width=8cm]{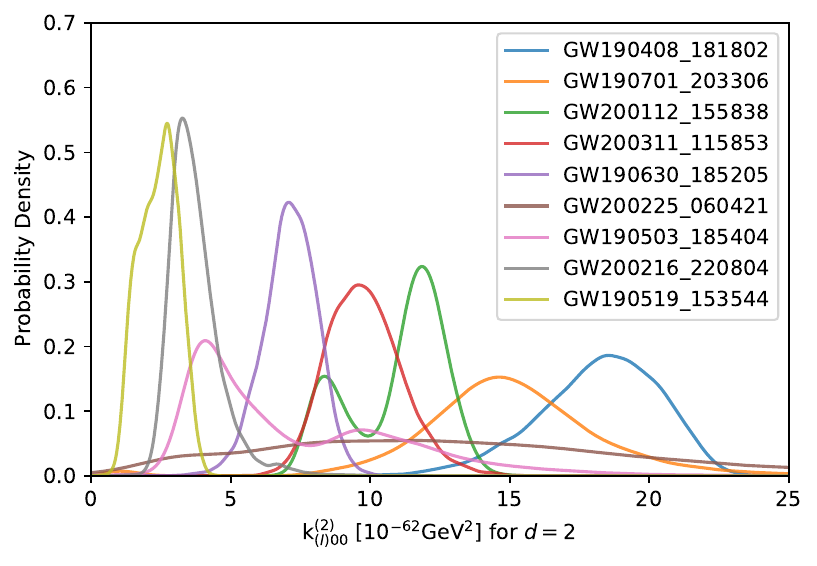}
\caption{The posterior distributions for $ k_{(I)00}^{(2)}$ from 9 excluded GW events (the excluded events are listed in Table.~\ref{excluded_list}). 
\label{kI00_excluded}}
\end{figure}

For the mass dimension $d=2$ case, the phase corrections (with $\delta\Psi_1=0$) in the modified waveform in (\ref{hplus}) and (\ref{hcross}) is expressed
\bqn
\delta \Psi_2 = A_{\bar \mu}(\pi f)^{-1},
\eqn
where
\bqn
A_{\bar \mu}= - \frac{1}{4 \sqrt{\pi}  } k_{(I)00}^{(2)} \int_0^z \frac{(1+z')^{-2}}{H_0\sqrt{\Omega_m(1+z')^{3}+ \Omega_\Lambda}}dz'.\nb\\
\eqn
Here we adopt a Planck cosmology with $\Omega_m=0.315$, $\Omega_\Lambda=0.685$, and $H_0 ={1.44 \times 10^{-42}\; {\rm GeV}}$ \cite{Planck:2018vyg}. The parameter $A_{\bar \mu}$ is the parameter we sampled in the Bayesian inference along with other GR parameters. We use the uniform prior for parameter $A_{\bar \mu}$ in our analysis. Then from the marginal posterior distributions of $A_{\bar \mu}$ and the redshift $z$ of the analyzed GW events, one can obtain posterior distributions of $k_{(I)00}^{(2)}$. In Fig.~\ref{kI00} we display the marginalized posterior distributions of $k_{(I)00}^{(2)}$
from selected GW events in the GWTC-3. For most GW events we analyze in each test, we do not find any significant signatures of Lorentz and diffeomorphism violation due to the coefficient $k_{(I)00}^{(2)}$. 
A few events that suggest nonzero values for the non-GR coefficients $ A_{\bar \mu} $ are excluded from our analysis due to their contradiction with GR. The posterior posterior distributions for $k_{(I)00}^{(2)}$ from the excluded GW events are presented in Fig.~\ref{kI00_excluded}. It is mentioned in \cite{Wang:2017igw, Wang:2021gqm} that these results may arise from limitations in current waveform approximants, such as systematic errors during the merger phase or unaccounted physical effects like eccentricity. Consequently, we have excluded these events from our analysis. Table \ref{excluded_list} presents the list of excluded events along with the estimated constraints on $ k^{(2)}_{(I)00}$ for each excluded event.

In addition, we consider the coefficient $k_{(I)00}^{(2)}$ as a universal parameter for all GW events, then we obtain its combined constraint by multiplying the posterior distributions of the individual events together, which gives
\bqn
- 0.5 \times 10^{-63} \; {\rm GeV}^2 <k_{(I)00}^{(2)}<1.3 \times 10^{-63} \; {\rm GeV}^2\nb\\
\eqn
at 90\% confidence level. The lower and upper limits of $k_{(I)00}^{(2)}$ are represented by the vertical dash line in Fig.~\ref{kI00}. 

\subsection{$d=3$}

For the mass dimension $d=3$, the phase correction (with $\delta \Psi_2=0$) in the modified waveform in (\ref{hplus}) and (\ref{hcross}) takes the form
\bqn
\delta \Psi_1 = A_{\mu}\ln u,
\eqn
with 
\bqn
A_{\mu} &=& \left(\sum_{jm} Y_{jm}({\bf \hat n})k_{(V)jm}^{(3)} \right) \nb\\
&& ~~~~ \times \int_0^z \frac{(1+z')^{-1}}{H_0\sqrt{\Omega_m(1+z')^{3}+ \Omega_\Lambda}}dz'. \lb{Auu}
\eqn
Here $A_{\mu}$ is the parameter we sampled in the Bayesian inference along with other GR parameters. We use the uniform prior for parameter $A_{\mu}$ in our analysis. As we mentioned, for mass dimension $d=3$, the coefficients $k_{(V)jm}^{(3)}$ have four independent components, $k_{(V)00}^{(3)}$, $k_{(V)10}^{(3)}$, ${\rm Re}k_{(V)11}^{(3)}$, and ${\rm Im}k_{(V)11}^{(3)}$. These components are entirely tangled together. 
Here we adopt an approach by using the ``maximum-reach" method, with which one can constrain each of these components separately \cite{Shao:2020shv, Wang:2021ctl, Niu:2022yhr}. This implies that when one considers one of these components, the others are set to zero. 

\begin{figure*}
\centering
\includegraphics[width=8cm]{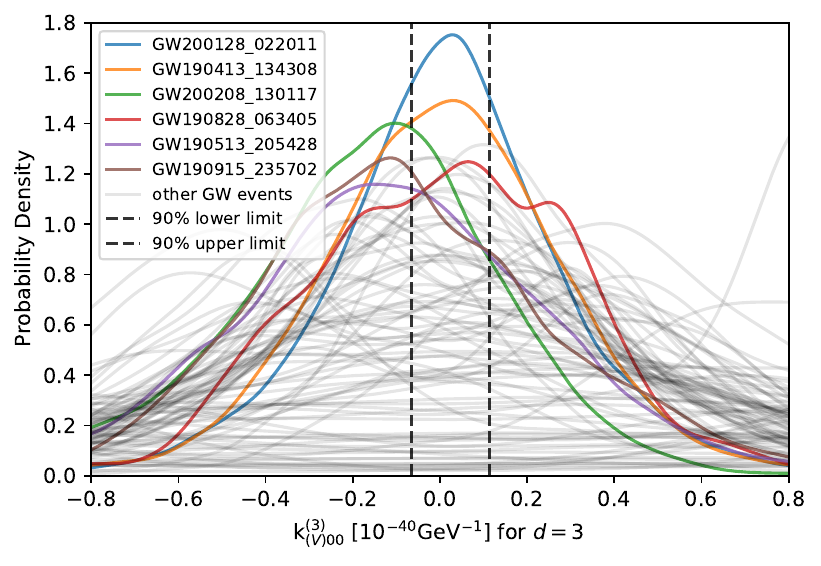}
\includegraphics[width=8cm]{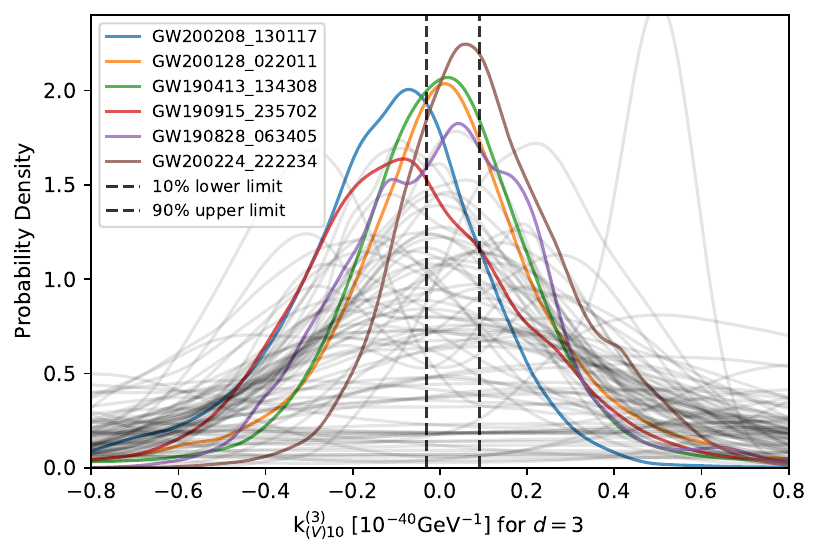}
\includegraphics[width=8cm]{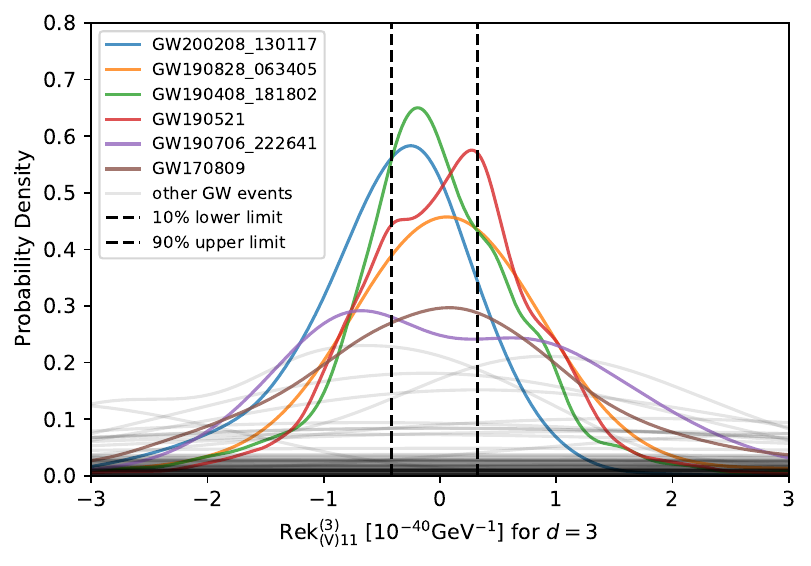}
\includegraphics[width=8cm]{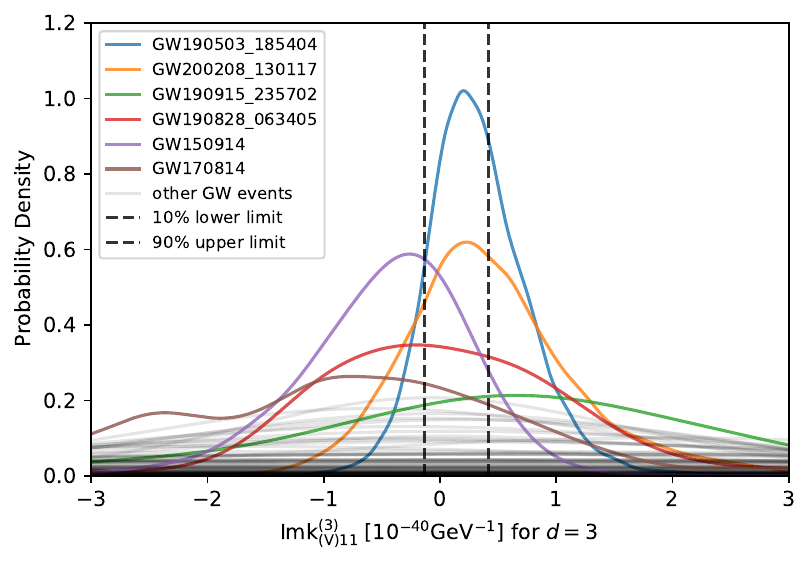}
\caption{The posterior distributions for $ k_{(V)00}^{(3)}$, $ k_{(V)10}^{(3)}$, ${\rm Re} k_{(V)11}^{(3)}$, and ${\rm Im} k_{(V)11}^{(3)}$ from selected GW events in the LIGO-Virgo-KAGRA catalog GWTC-3. The legend indicates the events that give the six tightest constraints. The vertical dash lines denote the 90\% intervals for the coefficients $ k_{(V)00}^{(3)}$, $ k_{(V)10}^{(3)}$, ${\rm Re} k_{(V)11}^{(3)}$ from combined results. 
\label{kVij}}
\end{figure*}

Then the posterior samples of each component of $ k_{(V)jm}^{(3)}$ can be calculated from the marginal posterior distributions of $ A_\mu$, right ascension (ra), declination (dec), and redshift $z$ of the analyzed GW events, via Eqs. (\ref{Auu}). Fig.~\ref{kVij} presents the marginalized posterior distributions of $ k_{(V)00}^{(3)}$, $ k_{(V)10}^{(3)}$, ${\rm Re}k_{(V)11}^{(3)}$, and ${\rm Im} k_{(V)11}^{(3)}$ from selected GW events in the GWTC-3. The lower and upper limits of each coefficient $ k_{(V)00}^{(3)}$, $ k_{(V)10}^{(3)}$, ${\rm Re}k_{(V)11}^{(3)}$, and ${\rm Im} k_{(V)11}^{(3)}$. $ k_{(I)jm}^{(6)}$ are represented by the vertical dash line in each figure of Fig.~\ref{kVij}. For most GW events we analyze in each test, we do not find any significant signatures of Lorentz and diffeomorphism violation due to the coefficient $ k_{(V)ij}^{(3)}$. In addition, for each coefficient of $k_{(V)00}^{(3)}$, $k_{(V)10}^{(3)}$, ${\rm Re} k_{(V)11}^{(3)}$, and ${\rm Im} k_{(V)11}^{(3)}$, we consider each of them as a universal parameter for all GW events, and then we obtain their combined constraints separately by multiplying the posterior distributions of the individual events together. Table.~\ref{kvij2} summarizes the $90\%$ confidence interval of each coefficients $ k_{(V)00}^{(3)}$, $ k_{(V)10}^{(3)}$, ${\rm Re} k_{(V)11}^{(3)}$, and ${\rm Im} k_{(V)11}^{(3)}$. From both the Fig.~\ref{kVij} and Table.~\ref{kvij2}, it is obvious that the posterior samples and the $90\%$ confidence interval of each coefficient $k_{(V)ij}^{(3)}$ are all consistent with zero, which indicates  {there are no} signatures of the Lorentz and diffeomorphism violations arising in the linearized gravity of SME has been found in the GW signals.

\begin{table}
\caption{The list of the excluded events in the analysis for $d=2$ case and their constraints on $k_{(I)00}^{(2)}$ at 90\% confidence interval. It is obvious that these constraints favor non-zero values of the non-GR coefficient $k_{(I)00}^{(2)}$.}
\label{excluded_list}
\centering  
\begin{ruledtabular}
    \begin{tabular}{c|cc}
      Coefficient & Events  &  Constraint ($10^{-62} \; {\rm Gev}^{2}$)\\
    \hline
     & $ \rm GW190408\_181802$ & $(15.2, 20.9)$\\
     & $\rm GW190503\_185404$ & $(3.5, 12.6)$\\
     &    $\rm GW190519\_153544$ & $(1.5, 3.3)$\\
     &   $\rm GW190630\_185205$ & $(5.9, 8.3)$ \\
     $k_{(I)00}^{(2)}$ &   $\rm GW190701\_203306$ & $(11.5, 19.1)$ \\
     &   $\rm GW200112\_155838$ & $(8.2, 12.9)$ \\
     &   $\rm GW200216\_220804$ & $(2.7, 4.9)$ \\
     &   $\rm GW200225\_060421$ & $(4.2, 23.5)$ \\
     &   $\rm GW200311\_115853$ & $(8.1,11.6)$ \\
    \end{tabular}
    \end{ruledtabular}
\end{table}

\begin{table}
\caption{$90\%$ confidence interval of each component of the Lorentz- and  diffeomorphism-violating coefficients $k_{(V)jm}^{(3)}$ from 90 GW events in the GWTC-3 catalog.}
\label{kvij2}
\centering  
\begin{ruledtabular}
    \begin{tabular}{cc|cc}
    $j$ & $m$ & Coefficient  &  Constraint ($10^{-41} \; {\rm Gev}^{-1}$)\\
    \hline
    $0$ & $0$ & $k^{(3)}_{(V)00}$ & $(-0.65,1.13)$\\
    $1$ & $0$ & $k^{(3)}_{(V)10}$ & $(-0.3,0.9)$\\
        & $1$ & Re   $k^{(3)}_{(V)11}$ & $(-4.2,3.2)$\\
        &     & Im   $k^{(3)}_{(V)11}$ & $(-1.2,4.2)$\\
    \end{tabular}
    \end{ruledtabular}
\end{table}

\section{conclusion}
\renewcommand{\theequation}{6.\arabic{equation}} \setcounter{equation}{0}
\lb{sec6}

The detection of GW signals by the LIGO-Virgo-KAGRA Collaboration marked the beginning of a new era in testing gravity in the strong-field regime. In this study, we investigate the effects of Lorentz- and diffeomorphism-violating effects in the linearized gravity on GW propagation within the framework of SME. Using an approach similar to that in the photon sector of the SME \cite{Kostelecky:2009zp}, we derive a modified dispersion relation for GWs with the Lorentz- and -violating effects, which lead to anisotropy, birefringence, and dispersion effects in the propagation of gravitational waves. With these modified dispersion relations, we then calculate the dephasing effects due to the Lorentz and diffeomorphism violations in the waveforms of gravitational waves produced by the coalescence of compact binaries.

With the distorted waveforms, one can test the Lorentz and diffeomorphism invariance of gravity by comparing the modified waveform with the GW strain data from the GW detectors. Several previous works have been carried out for testing Lorentz symmetry in Lorentz-violating linearized gravity for mass dimension $ d \geq 4$ cases. This study explores the effects of diffeomorphism violations induced by Lorentz-violating terms with mass dimensions $ d=2$ and $ d=3$ respectively. Using the SME framework for Lorentz-violating linearized gravity, we perform Bayesian inference with modified waveforms incorporating diffeomorphism-violating effects on GW events from the GWTC-3 catalog. We find no significant evidence of diffeomorphism violations in most GW data and provide 90\% confidence intervals for each diffeomorphism-
violating coefficients. 

Our results, illustrated in Fig. \ref{kI00} for $ d=2$ case, and Fig.~\ref{kVij} and Table.~\ref{kvij2} for $ d=3$ case, show no evidence of Lorentz and diffeomorphism violations. Accordingly, we report constraints on the coefficients $ k_{(I)00}^{(2)}$ describing anisotropic non-birefringent effects for $ d=2$ and coeffcients $ k_{(V)00}^{(3)}$, $ k_{(V)10}^{(3)}$, ${\rm Re} k_{(V)11}^{(3)}$, and ${\rm Im} k_{(V)11}^{(3)}$ describing birefringent anisotropic effects for $ d=3$. Nevertheless, the medians of all components remain near zero and thus are consistent with the results of GR prediction. Looking to the future, the next generation of GW detectors, with enhanced sensitivity and the ability to observe lighter and more distant binary black hole (BBH) and binary neutron star (BNS) systems, is anticipated to tighten further constraints on Lorentz and diffeomorphism violations induced dispersions in GW propagation.

\section*{Acknowledgements}

We appreciate the helpful discussions with Dr. Xiao Zhi, Bo-Yang Zhang, and Yuan-Zhu Wang. This work is supported by the National Key Research and Development Program of China under Grant No. 2020YFC2201503, the National Natural Science Foundation of China under Grants No.~12275238 and No.~11675143, the Zhejiang Provincial Natural Science Foundation of China under Grants No.~LR21A050001 and No.~LY20A050002, and the Fundamental Research Funds for the Provincial Universities of Zhejiang in China under Grant No.~RF-A2019015. W.Z. is supported by the National Natural Science Foundation of China (Grants No. 12325301 and 12273035) and the National Key Research and Development Program of China (Grant No. 2022YFC2204602 and 2021YFC2203102), the Fundamental Research Funds for the Central Universities. 

This research has made use of data or software obtained from the Gravitational Wave Open Science Center (gwosc.org), a service of the LIGO Scientific Collaboration, the Virgo Collaboration, and KAGRA. This material is based upon work supported by NSF's LIGO Laboratory which is a major facility fully funded by the National Science Foundation, as well as the Science and Technology Facilities Council (STFC) of the United Kingdom, the Max-Planck-Society (MPS), and the State of Niedersachsen/Germany for support of the construction of Advanced LIGO and construction and operation of the GEO600 detector. Additional support for Advanced LIGO was provided by the Australian Research Council. Virgo is funded, through the European Gravitational Observatory (EGO), by the French Centre National de Recherche Scientifique (CNRS), the Italian Istituto Nazionale di Fisica Nucleare (INFN) and the Dutch Nikhef, with contributions by institutions from Belgium, Germany, Greece, Hungary, Ireland, Japan, Monaco, Poland, Portugal, Spain. KAGRA is supported by the Ministry of Education, Culture, Sports, Science and Technology (MEXT), Japan Society for the Promotion of Science (JSPS) in Japan; National Research Foundation (NRF) and Ministry of Science and ICT (MSIT) in Korea; Academia Sinica (AS) and National Science and Technology Council (NSTC) in Taiwan.

The data analyses and results visualization in this work made use of \texttt{BILBY} \cite{Romero-Shaw:2020owr, Ashton:2018jfp}, \texttt{dynesty} \cite{Speagle:2019ivv}, \texttt{LALSuite} \cite{LALSuite}, \texttt{Numpy} \cite{Harris:2020xlr, vanderWalt:2011bqk}, \texttt{Scipy} \cite{Virtanen:2019joe}, and \texttt{matplotlib} \cite{Hunter:2007ouj}.

\appendix


%

\end{document}